\documentclass[atoms,review,submitted,moreauthors,10pt,a4paper]{mdpi1}
\firstpage{1}
\articlenumber{x}
\doinum{10.3390/------}
\pubvolume{xx}
\pubyear{2020}
\copyrightyear{2020}
\history{Received: date; Accepted: date; Published: date}

\Title{Photoionization of Astrophysically Relevant Atomic Ions at PIPE}

\Author{Stefan Schippers$^{1,\ast}$\orcidA{}, Alfred M\"uller$^2$\orcidB{}}
\AuthorNames{Stefan Schippers, Alfred Müller}

\address[1]{%
$^{1}$ \quad I. Physikalisches Institut, Justus-Liebig-Universität Gießen, Heinrich-Buff-Ring 16, 35392 Giessen, Germany; stefan.schippers@physik.uni-giessen.de\\
$^{2}$ \quad Institut f\"{u}r Atom- und Molek\"{u}lphysik, Justus-Liebig-Universit\"{a}t Gie{\ss}en, Leihgesterner Weg 217, 35392 Giessen, Germany
}
\corres{Correspondence: stefan.schippers@physik.uni-giessen.de}

\abstract{We review recent work on the photoionization of atomic ions of astrophysical interest that has been carried out at the photon-ion merged-beams setup PIPE, a permanently installed end station at the XUV beamline P04 of the PETRA\,III synchrotron radiation source operated by DESY in Hamburg, Germany. Our results on single and multiple $L$-shell photoionization of Fe$^+$, Fe$^{2+}$, and Fe$^{3+}$ ions and on single and multiple $K$-shell photoionization of C$^-$, C$^+$, C$^{4+}$, Ne$^+$, and Si$^{2+}$ ions are discussed in astrophysical contexts. Moreover, these experimental results bear witness of the fact, that the implementation of the photon-ion merged-beams method at one of the world's brightest synchrotron light sources has led to a breakthrough for the experimental study of atomic inner-shell photoionization processes with ions.}


\begin{document}


\section{Introduction}

Much of the baryonic matter in the Universe is in a plasma state. The interpretation of the astronomical observations of cosmic plasmas requires a quantitative understanding of the quantum processes that lead to the emission or absorption of photons and that govern the charge balance of atoms and ions  in a plasma. Photoionization and photoabsorption of atomic ions are important in connection with radiation transport, e.g., in stars \cite{Pain2017} or kilonovae \cite{Tanaka2020} and whenever a cosmic plasma is within the line of sight between the observer and a radiation source such as, e.g., a star, an x-ray binary, or an active galactic nucleus. Absorption spectra recorded by X-ray telescopes contain spectral signatures of the atomic ions contained in the plasma \cite{Paerels2003a}. The most prominent features are due to the cosmically most abundant elements. Next to hydrogen and helium these are C, N, O, Ne, Mg, Si, S, and Fe \cite{Asplund2009}. The astrophysically motivated atomic-data needs concerning these and other elements from the Periodic Table have been highlighted repeatedly (see, e.g., \cite{Ferland2003a,Kallman2007a,Savin2012,Smith2014,LynasGray2018,Smith2020}).

Here, we briefly review recent progress in experimental photoionization that has been accomplished using interacting photon and ion beams at one of the world's brightest 3rd generation synchrotron light sources. We present experimental cross sections for $L$-shell photoionization of Fe$^+$, Fe$^{2+}$ and Fe$^{3+}$ ions, and for $K$-shell photoionization of C$^-$, C$^+$, C$^{4+}$, Ne$^+$, and Si$^{2+}$ ions. In addition, we discuss the relevance of our results for the modeling of astrophysical plasmas.

\section{Experimental Technique}

Photoionization of ions by a single photon requires photon energies that exceed the ionization potential of the ion to be investigated. For stable atomic ions these range from 10 eV for Ba$^+$ to 132 keV for U$^{91+}$ \cite{Kramida2019}. Thus, photon energies from the vacuum ultraviolet (VUV) to the hard X-ray bands are needed for investigating photoionization of ions across the entire periodic table. Powerful laboratory sources for these types of radiation are hot plasmas and synchrotron light sources, which both have been used for photoabsorption and photoionization studies with atomic ions. The dual laser plasma (DLP) technique \cite{Kennedy2004a} uses a laser-generated hot plasma as a back-lighter for absorption measurements with ions in a second laser-generated plasma. In contrast to the broad spectral distribution of the radiation from a hot plasma, synchrotron radiation has a much narrower photon-energy bandwidth and is freely tunable over large energy ranges. Moreover, modern 3rd generation synchrotron light sources provide a high photon flux which is a prime necessity for experiments with dilute targets such as ions, whose mutual electrostatic repulsion entails low particle densities.  The density of  ionic targets can be increased in ion traps where the ion cloud can be compressed by external fields and its density can be increased by applying cooling techniques. Nevertheless, the signal rates from such arrangements are usually still rather low and, therefore, photoionization of trapped atomic ions has been performed  in only a few cases \cite{Kravis1991,Thissen2008a,Bizau2011,Simon2010a,Simon2010,Hirsch2012}, so far.

\begin{figure}[ttt]
\centering
\includegraphics[width=0.9\textwidth]{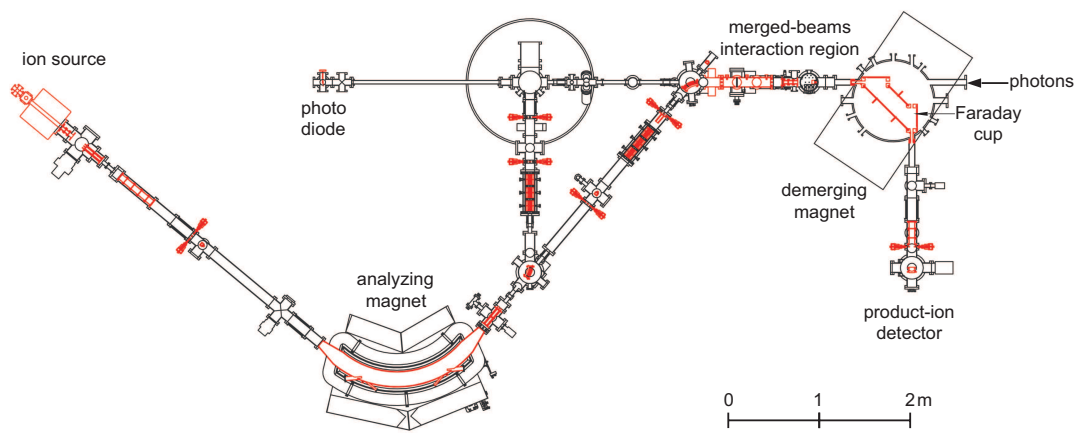}
\caption{\label{fig:PIPE} Sketch of the photon-ion merged-beams setup PIPE (\underline{P}hoton-\underline{I}on spectrometer at \underline{PE}TRA III). Ion-optical elements such as focussing lenses, beam deflectors, collimating slits and Faraday cup are drawn in red color. This figure is a differently labelled version of Figure~2 from \cite{Mueller2017} (reproduced by permission of the AAS).}
\end{figure}

The photon-ion merged-beams technique \cite{Lyon1986,Kjeldsen2006a,Schippers2016}  makes up for the diluteness of the ionic target by providing a large spatial overlap between the photon beam and the ion beam. Corresponding experimental setups were installed at several synchrotron light sources \cite{Lyon1986,Bizau1991,Oura1994,Kjeldsen1999b,Yamaoka2001,Covington2002,Schippers2014,Bizau2016a}. The most recent development is the use of a XUV laser for a photon-ion merged-beams experiment in a heavy-ion storage ring \cite{Lestinsky2016,Borovik2020}. Figure~\ref{fig:PIPE} sketches the photon-ion merged-beams setup PIPE \cite{Schippers2014} which is permanently installed at the variable polarization XUV beam line P04 \cite{Viefhaus2013} of the PETRA III synchrotron light source operated by DESY in Hamburg, Germany. Different types of ion sources can be mounted such that a large variety of ion beams can be produced. In the past years, experiments at PIPE have been performed with positive and negative atomic ions, small molecular ions, and endohedral fullerene ions \cite{Schippers2020}. After extraction from the ion source, which is operated on a potential of typically 6 kV, the ions are separated in the analyzing magnet according to their mass-to-charge ratio. Subsequently, the selected ion beam is brought to a coaxial overlap with the photon-beam by adjusting the electrostatic ion-optical elements of the ion beam line accordingly. The length of the overlap region is 1.7 m. The demerging magnet separates the more highly charged product ions from the primary ions. The demerging-magnet field strength is adjusted such that one selected product-ion charge state is directed onto a single-particle detector that counts the product ions, which hit the detector  with keV energies, with practically 100\% efficiency. The combination of a large interaction volume, a record-high photon flux (more than $10^{12}$~s$^{-1}$ at 0.01\% bandwidth across the entire 250--3000~eV P04 energy range), and a highly efficient and largely background-free product-detection scheme provides a world-unique sensitivity for photon-interaction studies with ionized matter in the gas phase.

\begin{figure}[ttt]
\centering
\includegraphics[width=0.7\textwidth]{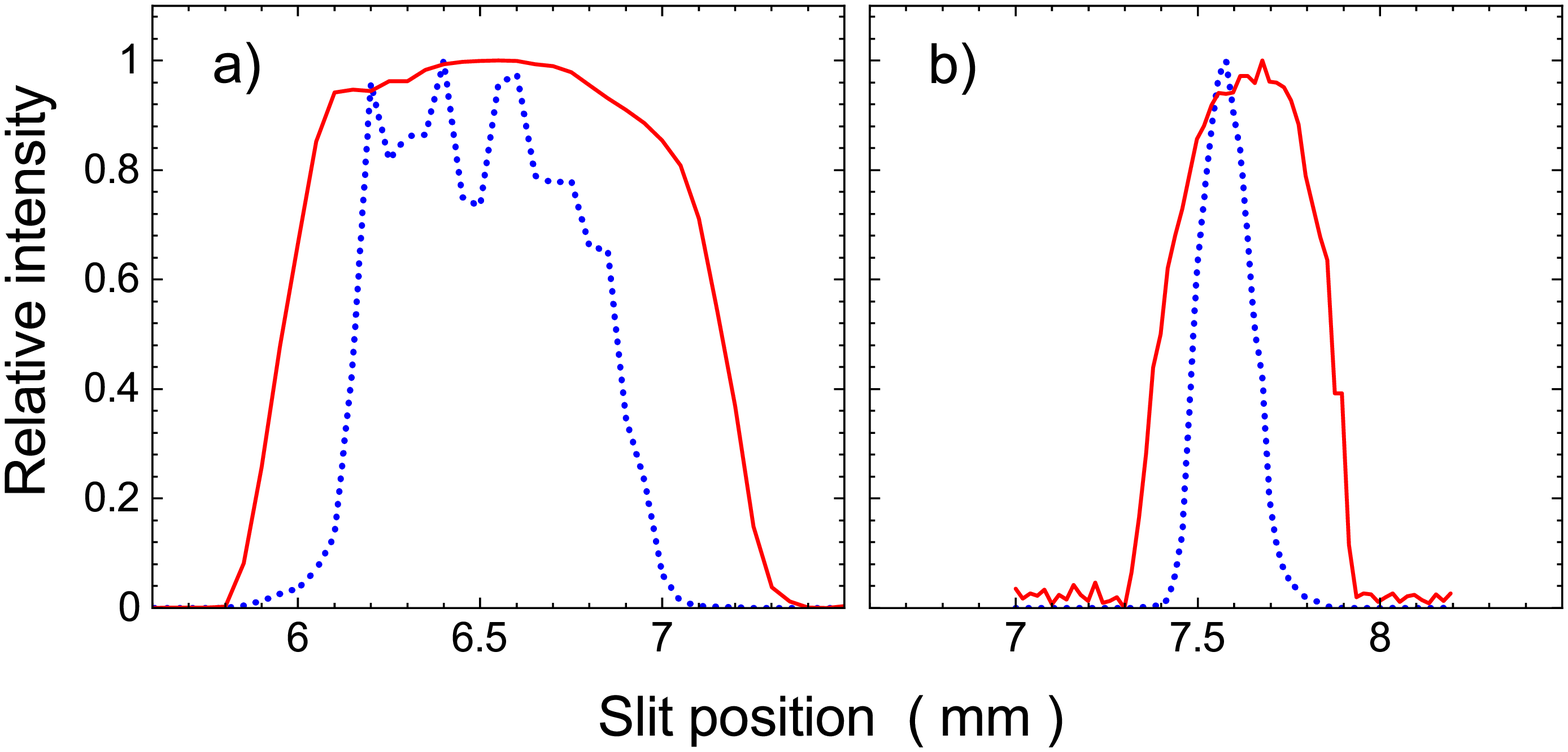}
\caption{\label{fig:slit} Measured horizontal (a) and vertical (b) beam profiles at the center of the interaction region \cite{Mueller2017} (reproduced by permission of the AAS). The red full and blue dashed curves represent the ion-beam (Ne$^+$) and photon-beam profiles, respectively.}
\end{figure}

An asset of the merged-beam technique, which is particularly important for applications in astrophysics, is its capability to provide \emph{absolute} photoionization cross sections. To this end, the photon flux, $\phi_\mathrm{ph}$, at PIPE is monitored with a calibrated photodiode and the primary ion current, $I_\mathrm{ion}$, is measured with a large Faraday cup that is located inside the demerging-magnet vacuum chamber (Fig.~\ref{fig:PIPE}). From these experimental quantities and from the measured product-ion count rate, R, the absolute cross section can be calculated as
\begin{equation}\label{eq:sigma}
 \sigma = R \frac{q\,e\,v_\mathrm{ion}}{\eta\,I_\mathrm{ion}\,\phi_\mathrm{ph}\,\mathcal{F}_L},
\end{equation}
where $qe$ and $v_\mathrm{ion}$ are the charge and the velocity of the primary ion and $\eta$ is the detection efficiency (usually $\eta=1.0$). The factor $\mathcal{F}_L$ quantifies the mutual spatial overlap of the ion beam and the photon beam (see \cite{Schippers2014} for details). It is determined by beam-profile measurements using slit scanners which probe the beam overlap at three different locations along the photon-ion interaction region. Fig.~\ref{fig:slit} shows  beam profiles measured at its center. This procedure results in a systematic uncertainty of the absolute cross-section scale of typically 15\% at a confidence level of 90\%  \cite{Schippers2014}.

The systematic uncertainty of the P04 photon-energy scale is typically $\pm$30--200~meV  after a calibration to absorption features in gases has been applied. As discussed in considerable detail in \cite{Mueller2017}, the remaining uncertainty is primarily due to the uncertainties of the calibration standards that are currently available in the soft X-ray range. Considering the ppm accuracies that can be obtained at hard X-ray beamlines with crystal monochromators, this situation for soft X-rays seems unsatisfying and calls for better calibration standards. As proposed in \cite{Mueller2018c} (see also below), few-electron ions could serve this purpose since the theoretical uncertainties of photoionization-resonance positions for such fundamental atomic systems are on the meV level if state-of-the-art atomic theory is applied (see, e.g., \cite{Yerokhin2017,Yerokhin2017a,Mueller2018c,Machado2020}).


\section{L-shell ionization of low-charged iron ions}

A particular astrophysical data need was addressed in a sequence of measurements with the low-charged iron ions Fe$^{+}$ \cite{Schippers2017}, Fe$^{2+}$ \cite{Schippers2020b}, and Fe$^{3+}$ \cite{Beerwerth2019}.  These ions and neutral iron atoms form the gaseous iron fraction in the interstellar medium \cite{Jensen2007a}. Another fraction of iron is chemically bound to interstellar dust grains. This fraction is an important parameter for tracing the evolution of the stellar mass distribution and, more generally, of the chemical evolution of the universe \cite{Jenkins2009}. The abundance of iron in the interstellar medium can be inferred from  astronomical X-ray observations of Fe $L$-shell features, and high-resolution data from X-ray satellites can be used to discriminate between the gaseous and solid forms of iron in the interstellar medium  \cite{Juett2006}. This requires laboratory data for $L$-shell absorption by solid iron compounds and by iron in the gas-phase. The available data for solids have been compiled in \cite{Miedema2013} and $2p$ absorption of neutral iron vapour was studied experimentally in \cite{Richter2004,Martins2006a}. However, data for $L$-shell photoionization of low-charged iron ions had not been available prior to our recent measurements at PIPE except for relative cross sections for single and double photoionization of Fe$^+$ \cite{Hirsch2012}. Further work on $L$-shell photoionization was only carried out for the higher charge states Fe$^{6+}$--Fe$^{10+}$ \cite{Blancard2018} and Fe$^{14+}$ \cite{Simon2010}. In addition, a number of merged beams experiments have been performed on valence shell photoionization of Fe$^+$ \cite{Kjeldsen2002c}, Fe$^{4+}$ \cite{Bizau2006a}, Fe$^{2+}$--Fe$^{6+}$ \cite{ElHassan2009}, and of Fe$^{3+}$, Fe$^{5+}$, and Fe$^{7+}$ \cite{Gharaibeh2011}.

\begin{figure}
\centering
\includegraphics[width=0.7\textwidth]{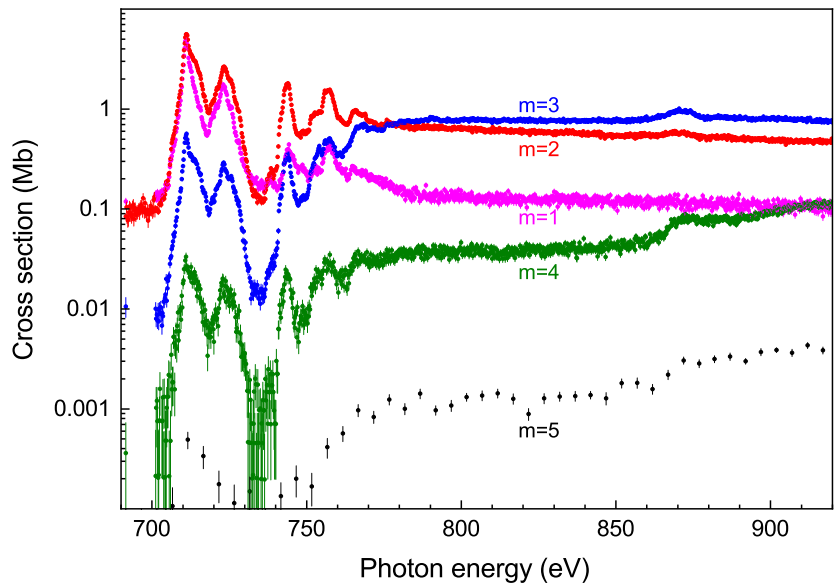}
\caption{\label{fig:Fe3} Experimental cross sections (1 Mb = $10^{-18}$~cm$^2$) for $m$-fold ionization of Fe$^{3+}$ ions in the energy range 690--920~eV that contains the thresholds for $2p$ and $2s$ ionization \cite{Beerwerth2019}. The observed resonance structures are associated with the excitation of a $2p$ or a $2s$ electron to a higher atomic subshell and subsequent autoionization.}
\end{figure}

\begin{figure}
\centering
\parbox[t]{0.41\textwidth}{\includegraphics[width=0.404\textwidth]{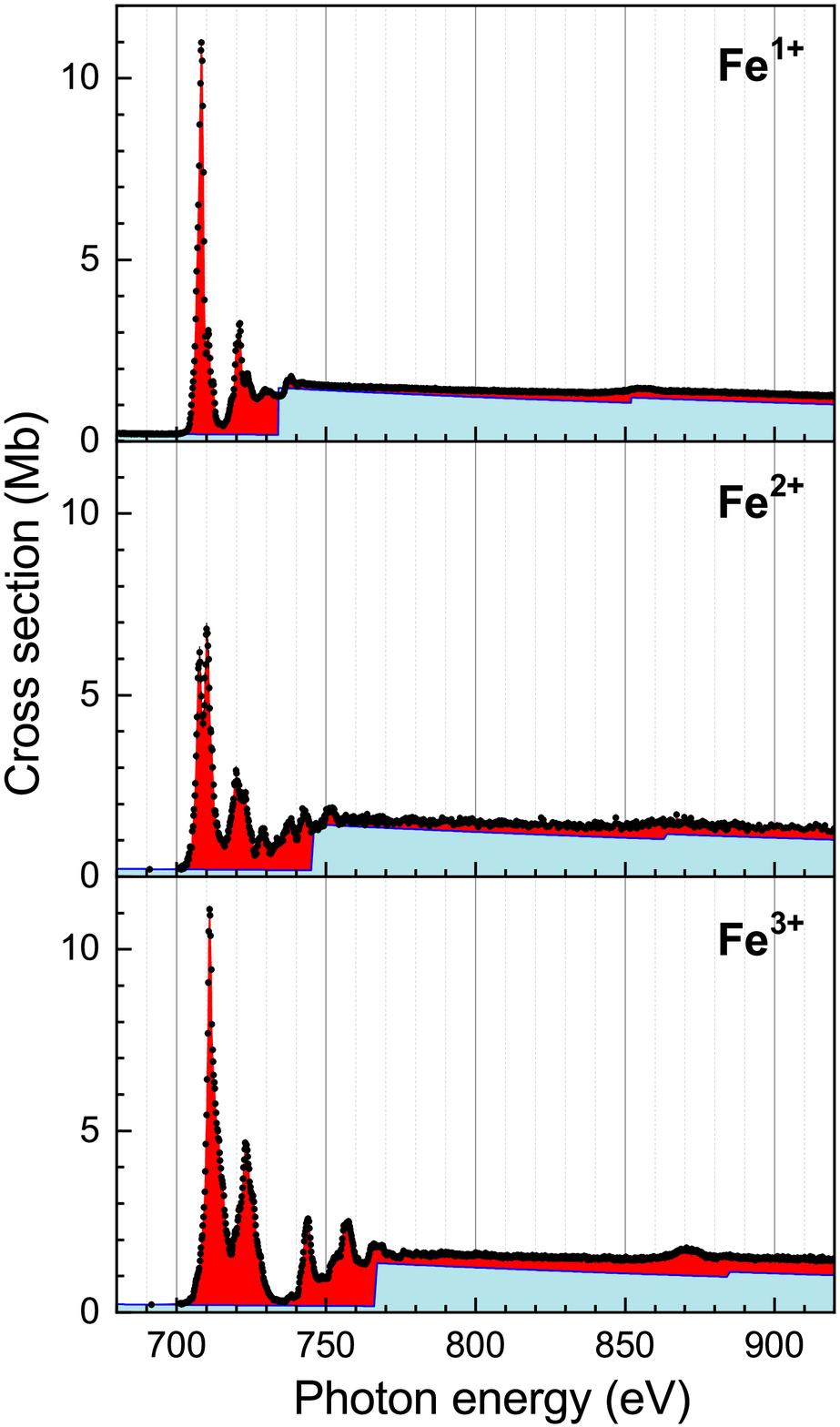}}\parbox{0.05\textwidth}{~}\parbox[t]{0.44\textwidth}{\includegraphics[width=0.44\textwidth]{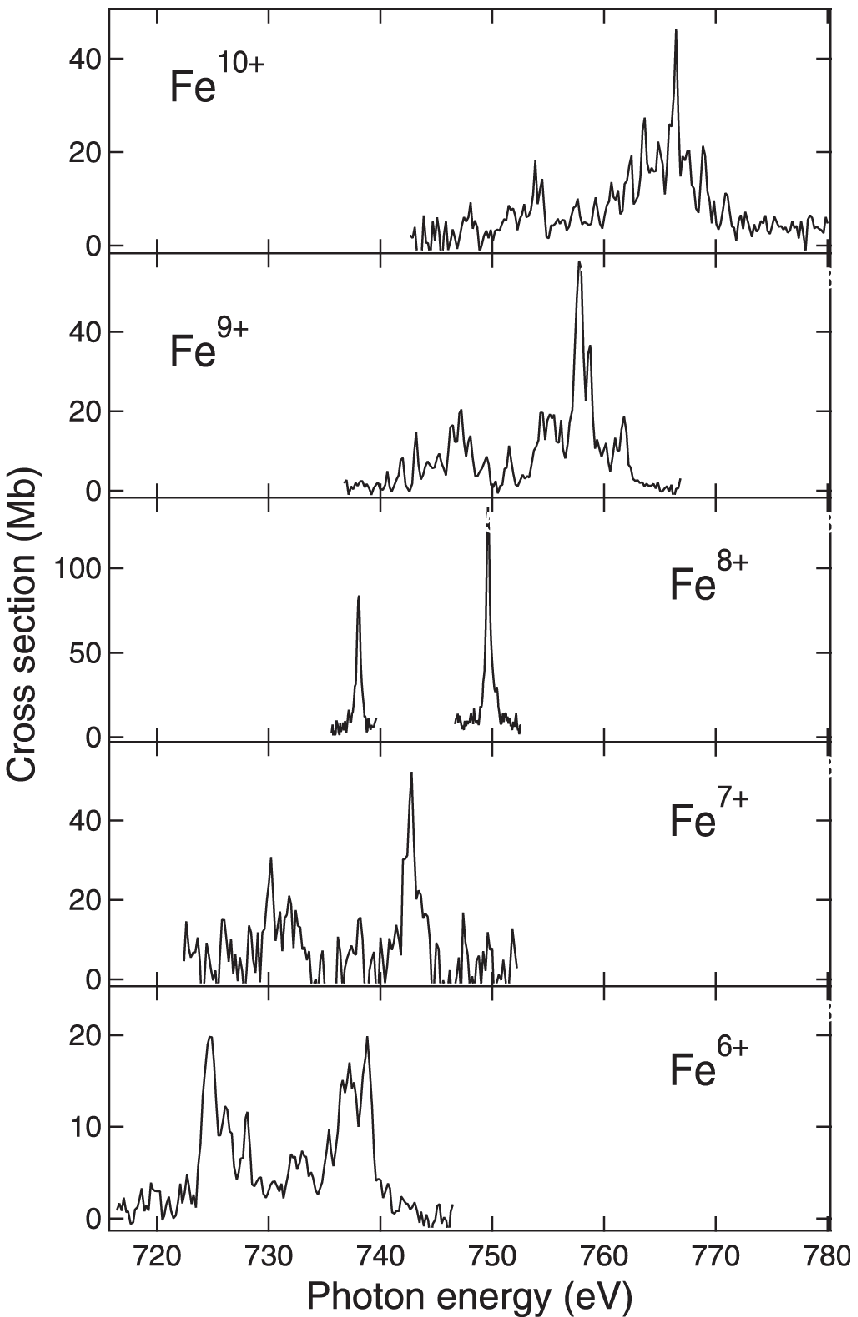}}
\caption{\label{fig:Fe123abs}Left panels: Experimental cross sections (symbols) from PIPE for photoabsorption of Fe$^+$ \cite{Schippers2017}, Fe$^{2+}$ \cite{Schippers2020b}, and Fe$^{3+}$ \cite{Beerwerth2019}. The uncertainty of the experimental energy scale amounts $\pm$0.2~eV. The light-shaded full curves are theoretical cross sections provided by Verner et al.~\cite{Verner1993a}. The steps in the theoretical cross sections occur at the computed thresholds for direct ionization of a $2p$ or a $2s$ electron. The photoionization resonances that dominate the experimental spectra below the $2p$ thresholds were not considered in the theoretical calculations.  Right panels: Experimental cross sections for single ionization of Fe$^{6+}$--Fe$^{10+}$ ions measured at SOLEIL \cite{Blancard2018} (reproduced by permission of the AAS).}
\end{figure}

Most of these previous studies considered only single ionization and in some cases also double ionization. At PIPE, cross sections could be measured for $m$-fold photoionization of Fe$^+$, Fe$^{2+}$, and Fe$^{3+}$ with $m$ ranging from 1 to 5 and in the case of Fe$^+$ \cite{Schippers2017} even to 6. As an example, Fig.~\ref{fig:Fe3} displays cross sections for up to five-fold ionization of Fe$^{3+}$ \cite{Beerwerth2019}. The cross-section scale spans several orders of magnitude ranging from several Mb to below one kb. Such low cross sections can only be accessed because of the high photon flux from the PETRA\,III synchrotron and because of the high selectivity and practically background-free detection of the product ions in the PIPE setup. All cross sections on display in Fig.~\ref{fig:Fe3} show the same resonance features, which are associated with the excitation of a $2p$ or a $2s$ electron to a higher $nl$ subshell. The two most prominent ones in the energy range 700--730 ~eV are due to $2p\to3d$ excitations. The separation of the two peaks corresponds to the $2p_{3/2}-2p_{1/2}$ spin-orbit splitting amounting to about 15~eV.  The resonances in the range 740--760~eV are dominated by $2p\to4d$ excitations according to atomic structure calculations that were carried out in support of the Fe$^{3+}$ experiment \cite{Beerwerth2019}. At energies below $\sim$770~eV, double ionization is the dominant channel. That energy corresponds to the threshold for direct $2p$ ionization. The theoretical value for the threshold energy is 766.9~eV~\cite{Verner1993a}. In contrast to $2p$ excitation, which occurs below this threshold and results in net double ionization, the $2p$ ionization process increases the ion charge state by one and, consequently, net triple ionization is the dominant ionization channel once direct ionization becomes energetically possible.

As can bee seen from Fig.~\ref{fig:Fe3}, the charge state distribution that results from inner-shell ionization is rather broad as compared to valence ionization. This is due to the multitude of deexcitation pathways that open up, once a $L$-shell hole is created. The deexcitation cascade involves radiative and autoionization transitions. The theoretical calculation of these cascades and the resulting product charge-state fractions is of considerable complexity. Already almost three decades ago, such calculations were performed  for astrophysically relevant ions, including iron, by Kaastra and Mewe \cite{Kaastra1993} who had to make simplifications to keep the computations tractable. Modern computers allow one to follow cascades in more detail and to include also many-electron processes such as autoionization accompanied by shake processes. This approach, which was taken in our supporting calculations for Fe$^+$--Fe$^{3+}$ \cite{Schippers2017,Schippers2020b, Beerwerth2019}, reproduced the experimental findings better than the earlier work of Kaastra and Mewe, but still leaves room for improvement. We just mention here that cascade calculations addressing $L$-shell ionization of Fe$^{2+}$ were carried out also by Ku\v{c}as et al.~\cite{Kucas2019,Kucas2020}. Their results are compared with our measured product charge-state fraction in \cite{Schippers2020b}.

Since all significant ionization processes have been measured one can very well approximate the Fe$^{3+}$ absorption cross section by the sum of all cross sections for $m$-fold ionization with $1\leq m \leq 5$. In Fig.~\ref{fig:Fe123abs} (left panels), the experimental absorption cross section for Fe$^+$--Fe$^{3+}$ \cite{Schippers2017,Schippers2020b, Beerwerth2019} are compared with the widely used theoretical results of Verner et al.~\cite{Verner1993a}. Their calculations comprised only  direct ionization processes and, therefore, do not reproduce the resonance features associated with $2p$ (and $2s$) excitation, which dominate the absorption cross sections below the $2p$ threshold and which are important for inferring  iron abundances from astronomical X-ray absorption spectra.  The resonance structures are significantly different for the different charge states such that a discrimination between these in the astronomical observations is feasible. As can be seen from the right panels of Fig.~\ref{fig:Fe123abs}, this is also true for the higher iron charge states. The displayed ionization cross sections for Fe$^{6+}$--Fe$^{10}$ \cite{Blancard2018} were measured at the MAIA setup \cite{Bizau2016a} at SOLEIL. They exhibit rather large statistical uncertainties as compared to the data from PIPE. This is due to the lower photon flux at SOLEIL as compared to PETRA\,III and also due to the fact that it is more difficult to obtain intense ion beams for more highly charged ions.

\section{K-shell ionization of light ions}

\begin{table}[b]
 \caption{\label{tab:Kshell} List of published experimental cross sections for $K$-shell photoionization of atomic ions with atomic numbers $Z\geq 2$ from photon-ion merged-beams experiments at the synchrotron light sources ALS (Berkeley, USA), ASTRID (Aarhus, Denmark), PETRA\,III (Hamburg, Germany), SOLEIL (Saint-Aubin, France), and SPring-8 (Hyogo, Japan). Columns  3 and 5 provide the experimental photon-energy range and the year of publication, respectively.}
\centering
 \begin{tabular}{rlclcl}
 \toprule
  $Z$ & Ion & Energy range (eV) &   Light source & Year & Reference \\
 \midrule
   2 & He$^+$ & \phantom{11}80 --  \phantom{1}140 & \textsc{astrid} & 2001 & \cite{Andersen2001b} \\
   2 & He$^-$ & \phantom{11}38 --  \phantom{11}44 & \textsc{als} & 2002 & \cite{Berrah2002} \\
   2 & He$^-$ & \phantom{11}43 --  \phantom{11}44 & \textsc{als} & 2004 & \cite{Bilodeau2004a} \\
   3 & Li$^-$ &  \phantom{11}56 --  \phantom{11}70 &  \textsc{astrid} & 2001 &\cite{Kjeldsen2001a} \\
   3 & Li$^-$ &  \phantom{11}56 --  \phantom{11}66 &  \textsc{als} & 2001 & \cite{Berrah2001} \\
   3 & Li$^+$ & \phantom{1}149 -- \phantom{1}181  & \textsc{als} & 2006 & \cite{Scully2006a} \\
   5 & B$^-$     & \phantom{1}187 -- \phantom{1}196   & \textsc{als}       & 2007 & \cite{Berrah2007a}   \\
  5 & B$^+$     & \phantom{1}193 -- \phantom{1}210   & \textsc{als}       & 2014 & \cite{Mueller2014a} \\
  5 & B$^{2+}$  & \phantom{1}195 -- \phantom{1}235   & \textsc{als}       & 2010 & \cite{Mueller2010} \\
  6 & C$^-$     & \phantom{1}280 -- \phantom{1}285   & \textsc{als}       & 2003 & \cite{Gibson2003a} \\
  6 & C$^-$     & \phantom{1}281 -- \phantom{1}282   & \textsc{als}       & 2006 & \cite{Walter2006a} \\
  6 & C$^-$     & \phantom{1}282 -- 1000             & \textsc{petra iii} & 2020 & \cite{Perry-Sassmannshausen2020}  \\
  6 & C$^+$     & \phantom{1}287 -- \phantom{1}290   & \textsc{als}       & 2004 & \cite{Schlachter2004a}\\
  6 & C$^+$     & \phantom{1}286 -- \phantom{1}326   & \textsc{petra iii} & 2015 & \cite{Mueller2015a}\\
  6 & C$^+$     & \phantom{1}286 -- \phantom{1}326   & \textsc{petra iii} & 2018 & \cite{Mueller2018}\\
  6 & C$^{2+}$  & \phantom{1}292 -- \phantom{1}323   & \textsc{als}       & 2005 & \cite{Scully2005b} \\
  6 & C$^{3+}$  & \phantom{1}300 -- \phantom{1}338   & \textsc{als}       & 2009 & \cite{Mueller2009a} \\
  6 & C$^{4+}$  & \phantom{1}358 -- \phantom{1}439   & \textsc{petra iii} & 2018 & \cite{Mueller2018c} \\
  7 & N$^+$     & \phantom{1}399 -- \phantom{1}406   & \textsc{soleil}    & 2011 & \cite{Gharaibeh2011a}\\
  7 & N$^+$     & \phantom{1}390 -- \phantom{1}435   & \textsc{petra iii} & 2019 & \cite{Bari2019} \\
  7 & N$^+$     & \phantom{1}415 -- \phantom{1}440   & \textsc{soleil}    & 2020 & \cite{McLaughlin2020}\\
  7 & N$^{2+}$  & \phantom{1}404 -- \phantom{1}442   & \textsc{soleil}    & 2014 & \cite{Gharaibeh2014}\\
  7 & N$^{3+}$  & \phantom{1}412 -- \phantom{1}414   & \textsc{soleil}    & 2013 & \cite{AlShorman2013}\\
  7 & N$^{4+}$  & \phantom{1}421 -- \phantom{1}460   & \textsc{soleil}    & 2013 & \cite{AlShorman2013}\\
  8 & O$^{-}$   & \phantom{1}526 -- \phantom{1}536   & \textsc{als}       & 2012 & \cite{Gibson2012}\\
  8 & O$^{-}$   & \phantom{1}524 -- \phantom{1}543   & \textsc{petra iii} & 2016 & \cite{Schippers2016a} \\
  8 & O$^+$     & \phantom{1}525 -- \phantom{1}540   & \textsc{spring-8}  & 2002 & \cite{Kawatsura2002a}\\
  8 & O$^{+}$   & \phantom{1}526 -- \phantom{1}620   & \textsc{soleil}    & 2015 & \cite{Bizau2015}\\
  8 & O$^{2+}$  & \phantom{1}526 -- \phantom{1}620   & \textsc{soleil}    & 2015 & \cite{Bizau2015}\\
  8 & O$^{3+}$  & \phantom{1}540 -- \phantom{1}600   & \textsc{soleil}    & 2014 & \cite{McLaughlin2014}\\
  8 & O$^{4+}$  & \phantom{1}550 -- \phantom{1}670   & \textsc{soleil}    & 2017 & \cite{McLaughlin2017} \\
  8 & O$^{5+}$  & \phantom{1}561 -- \phantom{1}570   & \textsc{soleil}    & 2017 & \cite{McLaughlin2017}\\
  9 & F$^{-}$   & \phantom{1}660 -- 1000             & \textsc{petra iii} & 2018 & \cite{Mueller2018b}\\
 10 & Ne$^+$    & \phantom{1}841 -- \phantom{1}858   & \textsc{spring-8}  & 2001 & \cite{Yamaoka2001}\\
 10 & Ne$^+$    & \phantom{1}840 -- \phantom{1}925   & \textsc{petra iii} & 2017 & \cite{Mueller2017}\\
 10 & Ne$^{2+}$ & \phantom{1}850 -- \phantom{1}863   & \textsc{spring-8}  & 2001 & \cite{Yamaoka2001}\\
 10 & Ne$^{3+}$ & \phantom{1}853 -- \phantom{1}873   & \textsc{spring-8}  & 2001 & \cite{Oura2001} \\
 14 & Si$^{2+}$ & 1830 -- 1880                       & \textsc{petra iii} & 2020 & \cite{Buhr2020}\\
\bottomrule
\end{tabular}
\end{table}

Deep inner-shell photoionization of atomic ions has been reviewed five years ago by M\"uller~\cite{Mueller2015}. This earlier review covers a large part of the data that were measured prior to the start of the experimental program at PIPE in 2013. Table~\ref{tab:Kshell} provides a comprehensive compilation of merged-beams studies on $K$-shell ionization of positive and negative atomic ions, which cover a time span of two decades. In the present review,  we concentrate on the recent results from PIPE for Ne$^+$, C$^-$, C$^+$, C$^{4+}$, and Si$^{2+}$ ions.

\subsection{$K$-shell photoionization of Ne$^+$}

\begin{figure}
\centering
\includegraphics[width=0.9\textwidth]{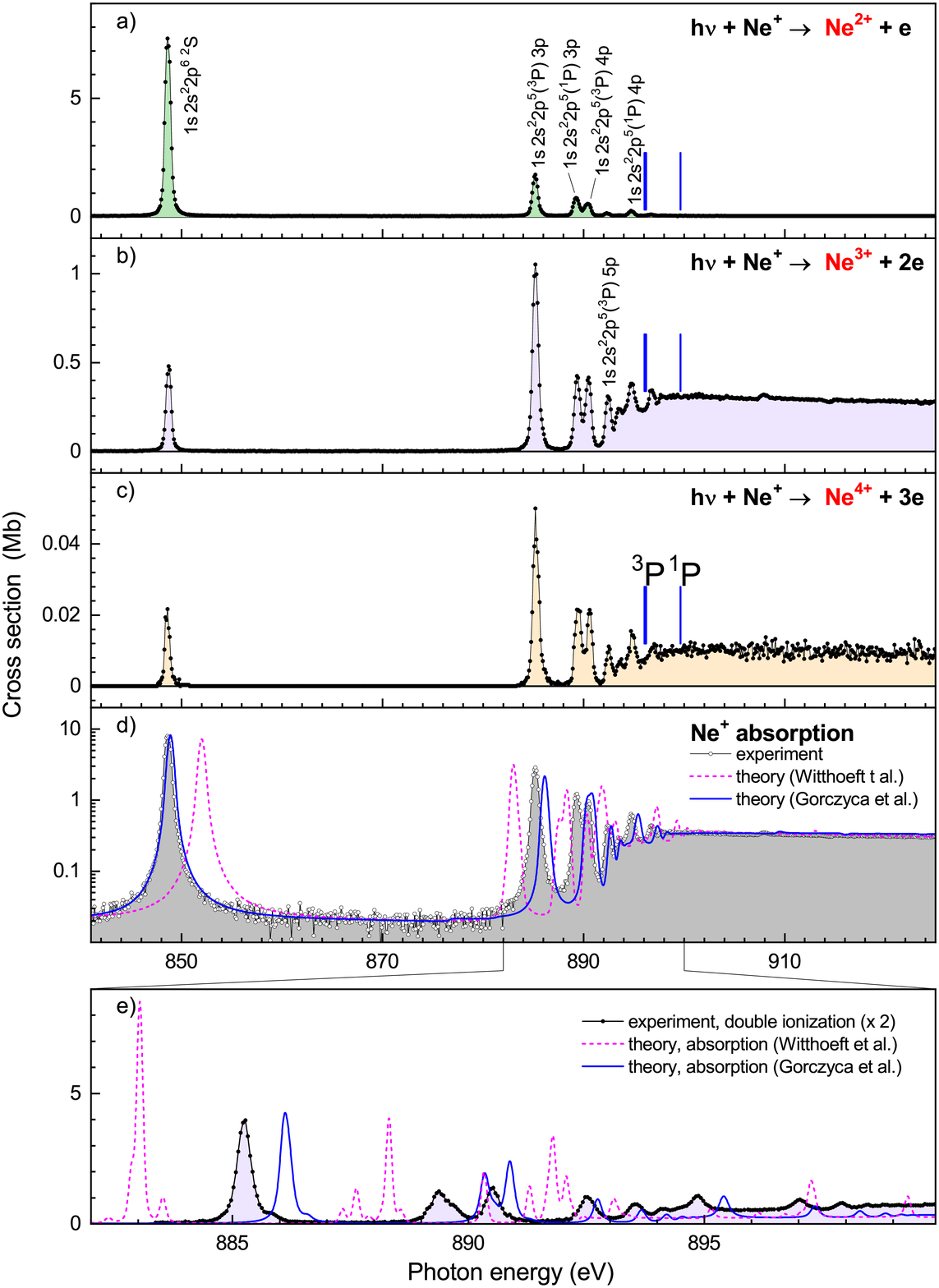}
\caption{\label{fig:Ne1} Cross sections for a) single ionization, b) double ionization, and c) triple ionization of Ne$^+$ ions \cite{Mueller2017} measured with a photon energy bandwidth of 500~meV. The experimental cross section in panel d) is the sum of the cross sections displayed in panels a), b), and c). It represents the Ne$^+$ absorption cross section, which is compared with the theoretical cross sections of Gorczyca et al.~\cite{Gorczyca2000a,Juett2006,Gatuzz2015} and Witthoeft et al.~\cite{Witthoeft2009}. Panel e) presents a comparison between the theoretical results with a high-resolution measurement (113 meV photon-energy bandwidth) of the cross section for double ionization (multiplied by a factor of 2) over a narrower energy range. For the comparisons, the theoretical cross sections were convolved with gaussians with full-widths-at-half-maximum of 500~meV in panel d) and 113 meV in panel e).}
\end{figure}

X-ray emission and absorption lines of singly charged neon are frequently encountered in the spectra recorded by X-ray observatories. They originate from a variety of cosmic objects such as stellar coronae, X-ray binaries, supernova remnants, the interstellar medium, galaxies, and active galactic nuclei \cite[][and references therein]{Liao2013}. Figure~\ref{fig:Ne1} shows experimental cross sections for single, double, and triple ionization of Ne$^+$ ions \cite{Mueller2017} that contain prominent resonance lines associated with the excitation of a $1s$ electron to a higher $np$ subshell. As discussed above already for iron, also here the absorption cross section can be well approximated by the cross-section sum over the different final ion charge states. In  Fig.~\ref{fig:Ne1}d, the experimental absorption cross section thus derived is compared to state-of-the-art theoretical results of Gorzcyca et al.~\cite{Gorczyca2000a,Juett2006,Gatuzz2015} and Witthoeft et al.~\cite{Witthoeft2009}. At energies above 900~eV, i.e, at energies well above the threshold for direct $K$-shell ionization, where the cross section in the displayed energy range is essentially flat, all results agree excellently with one another. However, there are significant discrepancies concerning the photoionization resonances.

As detailed in \cite{Mueller2017}, accurate resonance positions, resonance widths, and strength could be retrieved from the measured absolute cross sections. Even more detailed information on Auger transition rates and fluorescence yields could be obtained, in particular, for the $1s\,2s^2\,2p^6\;^2S_{1/2}\to 1s^2\,2s^2\,2p^5\;^2P_{1/2,\;3/2}$ $K\alpha$ transitions. As can be seen in Figs.~\ref{fig:Ne1}d and \ref{fig:Ne1}e, the theoretical resonance positions differ from the experimental ones by much more than the $\pm$0.2~eV uncertainty of the experimental energy scale. Moreover, there are also discrepancies concerning resonance widths and fine-structure splittings not only between experiment and theory but also between both state-of-the-art theoretical calculations.  This is most probably due to different computational approximations of the two different approaches, e.g.,  by choosing different, naturally limited sets of basis states in close-coupling calculations whose convergence is not always obvious. Clearly, experimental benchmarking is vitally needed for the accurate location of resonance lines and thus for their unambiguous identification in absorption and emission spectra from astronomical observations. One should also be aware of the fact that accurate line strengths are required for inferring elemental abundances from astrophysical line spectra. At the current level of theoretical accuracy, apparently also these quantities need to be checked against laboratory results.

\subsection{$K$-shell photoionization of carbon ions: C$^-$, C$^+$, C$^{4+}$}

\begin{figure}
\centering
\includegraphics[width=0.9\textwidth]{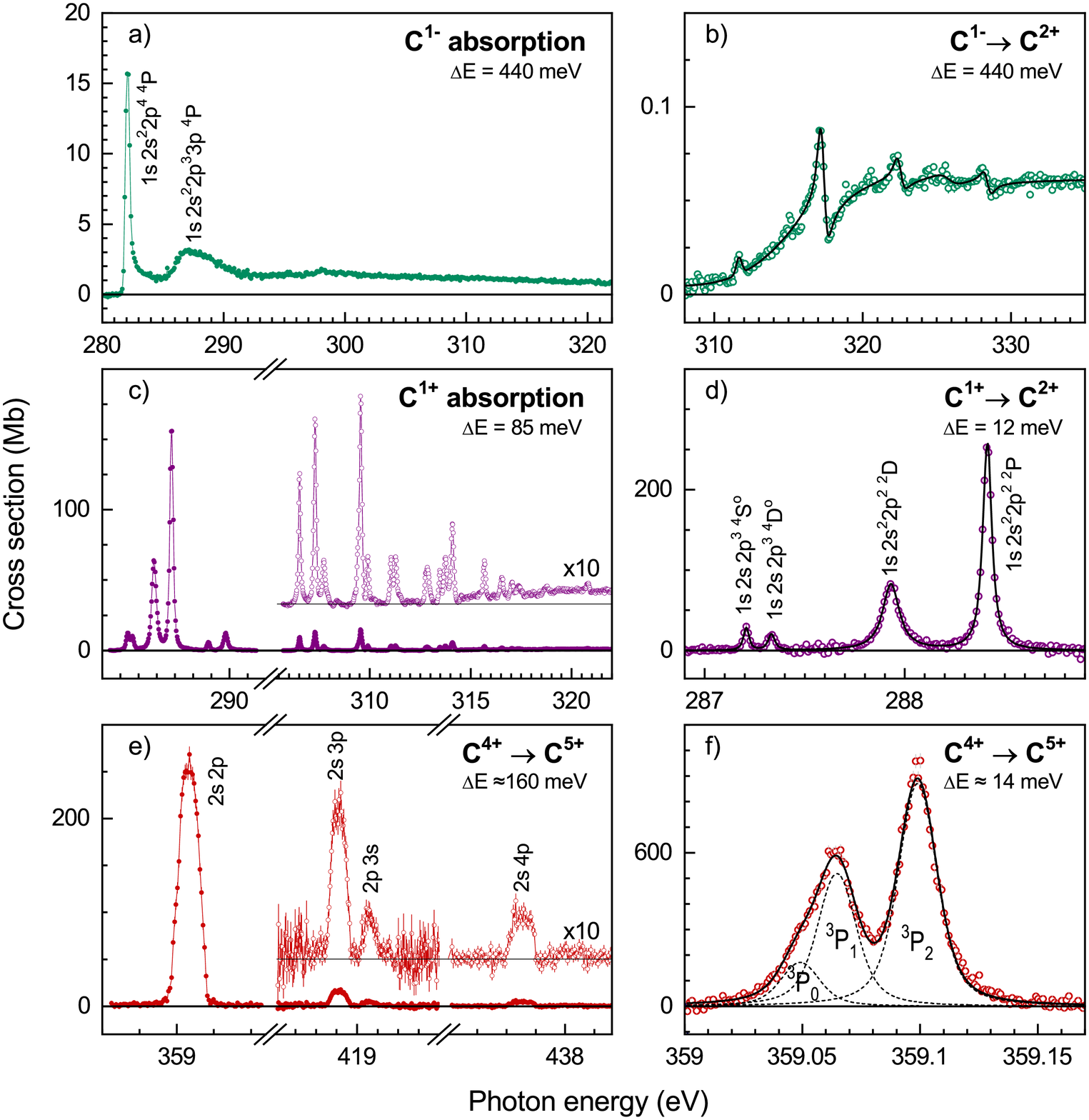}
\caption{\label{fig:Call} Experimental cross sections for photoabsorption and (multiple) photoionization of C$^-$ \cite{Perry-Sassmannshausen2020} \{panels a) and b)\}, C$^+$ \cite{Mueller2018} \{panels c) and d)\}, and C$^{4+}$ \cite{Mueller2018c} \{panels e) and f)\}. The C$^+$ and C$^{4+}$ ion beams were mixtures of ground-level and excited metastable ions (see text). For each measurement the experimental photon-energy bandwidth $\Delta E$ is specified. The left panels provide overviews, the right panels feature cross-section details that were measured partly with significantly smaller photon-energy bandwidths. For a better display of the data in panel b), a smooth cross section was subtracted representing the direct ionization of a $1s$ electron (see \cite{Perry-Sassmannshausen2020} for details). The full black lines in panels b), d), and f) result from resonance fits to the experimental data. The uncertainty of the experimental photon-energy scale is $\pm$200~meV for C$^-$, $\pm$30~meV for C$^+$, and (depending on photon energy)  $\pm$40~meV up to $\pm$50~meV for C$^{4+}$.}
\end{figure}

Carbon is the fourth most abundant element in the Solar system \cite{Asplund2009}.  High charge states of carbon such as C$^{4+}$ occur in many of the cosmic objects mentioned in the introduction. In our vicinity, C$^{4+}$ is found in  the solar wind, where its abundance has been shown to significantly depend on photoionization by the solar EUV and X-ray radiation \cite{Landi2015}. Lower-charged carbon, in particular C$^+$, plays a decisive role in the interstellar chemistry \cite{Larsson2012}. In fact, most of the molecules that have been detected in the interstellar medium to date are organic \cite{Tielens2013}. Some of these carbon containing molecules are negatively charged \cite{Millar2017}. However, the atomic ion C$^-$ has not been detected in any cosmic object. Potentially, this could be due to a lack of corresponding atomic data since the $K$-shell absorption cross section has become available for a wide range of photon energies only recently \cite{Perry-Sassmannshausen2020} (see  below). As can be seen from Tab.~\ref{tab:Kshell}, merged-beams experiments on $K$-shell ionization of  carbon ions were carried out at the ALS and at PETRA\,III (PIPE) covering the charge states -1,  1, 2, 3 and 4. In addition to this work on carbon $K$-shell ionization of atomic ions, there are also related studies with the carbon containing molecules CH$^+$~\cite{Mosnier2016}, La$_3$N@C$_{80}^+$~\cite{Hellhund2015},  and Sc$_3$N@C$_{80}^+$~\cite{Mueller2019} which are not discussed here.

Figure \ref{fig:Call} showcases some of the highlights obtained from photon-ion merged beams experiments with atomic carbon ions at PIPE. The displayed absorption and ionization cross sections for C$^-$, C$^+$, and C$^{4+}$ are distinctly different from one another. This  allows one to discriminate between these  ions in astronomical X-ray absorption spectra. The spectra of the positive ions are dominated by photoionization resonances of primary  ions being initially in the ground level or in a long-lived excited metastable level. The C$^+$ ion beam contained a 9:1 mixture of ions in the $1s^2\,2s^2\,2p\;^2P$ ground term and in the metastable $1s^2\,2s\,2p^2\;^4P$ term \cite{Mueller2018}. The C$^{4+}$ beam was a mixture of C$^{4+}$($1s^2\;^1S_0$) and C$^{4+}$($1s\,2s\;^3P_1$) ions with the metastable fraction amounting to 0.105 \cite{Mueller2018c}. In the C$^-$ beam, all ions were initially in the $1s^2\,2s^2\,2p^3\;^4S$ ground term. Unfortunately, there is no general method  for determining the fraction of metastable ions in an ion beam (for special cases see, e.g., \cite{Voulot2000,Covington2001a,Benis2018}).   For C$^+$, the ion-beam composition was inferred from a detailed comparison between measured and calculated absorption cross sections. This approach is somewhat dissatisfying since it relies on the accuracy of the theoretical methods, which is generally difficult to assess. In exceptional cases, one can relate the experimental photoionization cross  sections to experimental cross sections for the time inverse process of photorecombination by employing the principle of detailed balance \cite{Schippers2002b,Mueller2009a,Mueller2014a}. This method could be applied to  C$^{4+}$  using the  measured absolute cross section for photorecombination of C$^{5+}$ ions from a storage-ring experiment~\cite{Wolf1991}.

 The C$^-$ absorption cross section is displayed in Fig.~\ref{fig:Call}a.  Only two relatively weak  resonance features can be discerned. Generally, negative ions support less resonances than positive ions because the potential that binds the extra electron in a negative ion is comparatively shallow. Nevertheless, a number of hitherto mostly unknown resonances were discovered in the C$^-$ photoionization cross sections measured at PIPE as can been seen in Figs.~\ref{fig:Call}a and \ref{fig:Call}b. From a fundamental point of view, investigations of negative ions are interesting because their structure and dynamics is governed by strong correlation effects \cite{Andersen2004b}.  The $1s^2\,2s^2\,2p^3\;^4S\to 1s\,2s^2\,2p^4\;^4P$ resonance at about 282 eV (Fig.~\ref{fig:Call}a) was already experimentally studied earlier by Walter et al. \cite{Walter2006a}. These authors found that it can be well described by an asymmetric Breit-Wigner line shape which, in addition to the resonance itself,  accounts for the $K$-shell ionization threshold that occurs just 0.1~eV below the resonance energy. The broader resonance peaking at $\sim$287~eV (Fig.~\ref{fig:Call}a) is outside the experimental energy range of the earlier work. It has been tentatively assigned to  $1s^2\,2s^2\,2p^3\;^4S\to 1s\,2s^2\,2p^2\,3p\;^4P$ excitations \cite{Perry-Sassmannshausen2020}. The weak resonances at higher energies that are only visible in the experimental cross section for net triple ionization (Fig.~\ref{fig:Call}b)  could not be unambiguously  assigned. They exhibit asymmetric Fano line shapes caused by quantum mechanical interference of the resonance channel with  direct $1s+2s$ double ionization, which sets in at about 310~eV  \cite{Perry-Sassmannshausen2020}.  We just mention here that also double core-hole creation by direct $1s+1s$ ionization was observed in the five-fold ionization channel of the C$^-$ experiment.

Also for C$^+$ and C$^{4+}$, the most prominent resonances in the cross sections displayed in Figs.~\ref{fig:Call}c--\ref{fig:Call}f are associated with $1s\to np$ excitations with $n\geq 2$. The resonance structure is more complicated for C$^+$ than for C$^{4+}$ because of the larger number of open subshells and the correspondingly  larger number of fine-structure levels of  the excited C$^+$($1s\,2s^2\,2p\,np$) and C$^+$($1s\,2s\,2p^2\,np$) configurations as compared to the C$^{4+}$($2s\,np$) configurations. The latter are exclusively excited from the  metastable  C$^{4+}$($1s\,2s\;^3P_1$) ions. Their excitation from the C$^{4+}$($1s^2\;^1S_0$) ground level would be very much less efficient since this requires the simultaneous rearrangement of both electrons and the required photon energy would be almost 300~eV higher.

The careful evaluation of the measured data allows one to extract a considerable amount of detail concerning resonance parameters and deexcitation pathways.  For example, the measurement of the strengths and widths of the C$^+$($1s\,2s^2\,2p^2$) resonances in the single, double and threefold ionization channels provided unambiguous evidence and quantitative results for the elusive triple-Auger process, which is a manifestation of a genuine four-particle interaction \cite{Mueller2015a,Mueller2018}. The widths (and asymmetry parameters) of narrow resonances can be  extracted from high-resolution measurements  by fitting (Fano-)Voigt line profiles \cite{Schippers2018} to the measured data (Fig.~\ref{fig:Call}b,  \ref{fig:Call}d, and \ref{fig:Call}f).  Moreover, high photon-energy resolving powers $E/\Delta E$ facilitate the separation of individual fine-structure components as shown in Fig.~\ref{fig:Call}f for the C$^{4+}$($2s\,2p\;^3P$) resonance where $E/\Delta E = 25\,800$ was reached \cite{Mueller2018c}. This result is unique, since the fine-structure of resonance terms associated with deep inner-shell excitations has not yet been resolved in any other photoionization experiment with ions.

The original paper on the photoionization of C$^{4+}$ \cite{Mueller2018c} also comprises state-of-the-art theoretical calculations using relativistic many-body perturbation theory (RMBPT) \cite{Lindroth2012} with quantum electrodynamical (QED) additions \cite{Indelicato1987}. An in-depth discussion of these calculations is beyond the scope of the present review, but we want to point out that the C$^{4+}$ photoionization resonances  were calculated with an exceptionally high accuracy at an estimated uncertainty of the resonance energies of less than  $\pm$1~meV. At present, such an accuracy can only be achieved for few-electron ions where the description of  electron-electron--interaction effects is well under control. At the given level of theoretical uncertainty of resonance energies for He-like ions, these resonances can potentially be used to establish new  photon-energy calibration standards that are up to two orders of magnitude more accurate than what is currently available in the soft X-ray range, as is discussed in detail in  \cite{Mueller2017,Mueller2018c}.  Like many other technical applications, the field of X-ray astronomy will certainly  greatly benefit from such a development.

\subsection{$K$-shell ionization of silicon ions}

\begin{figure}
\centering
\includegraphics[width=0.5\textwidth]{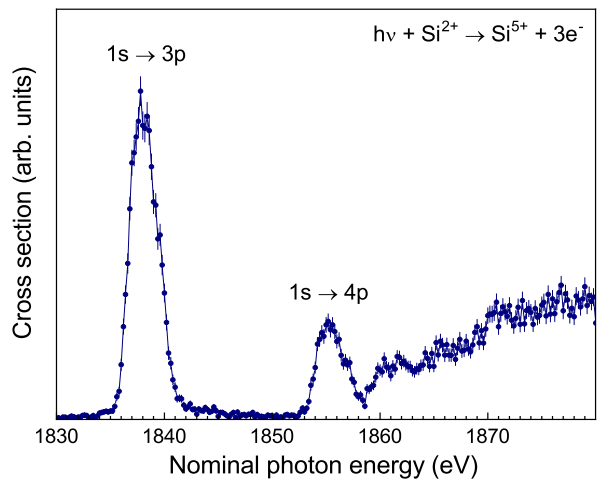}
\caption{\label{fig:Si2}Experimental cross section for $K$-shell triple ionization of Si$^{2+}$. These preliminary data from \cite{Buhr2020} are not on an absolute cross section scale, and the photon energy scale has not been calibrated, yet.}
\end{figure}

$K$-shell photoabsorption by silicon is used to trace its abundance in the interstellar medium where it is to a large part contained in dust particles  \cite{Zeegers2019} . Only a small fraction of the interstellar silicon is expected to be in the gas-phase. Recently, it has been suggested that gas phase absorption data for silicon ions are required  for an accurate modelling of the astronomically observed Si $K$-edge absorption features \cite{Schulz2016}. This has motivated theoretical work on the photoabsorption of silicon atoms and ions \cite{Witthoeft2009,Witthoeft2011,Kucas2012,Hasoglu2018}. At PIPE, we have recently measured cross sections for single and multiple photoionization of low-charged Si$^{+}$, Si$^{2+}$, ans Si$^{3+}$  ions. Figure \ref{fig:Si2} shows preliminary results for the $K$-shell photoionization of Si$^{2+}$([Ne]$\,3s^2\,^1S_0$) \cite{Buhr2020}. The experimental cross section, which has not yet been put on an absolute scale, has strong resonance contributions associated with $1s\to 3p$ and $1s\to 4p$ excitations. A more careful, currently ongoing analysis will reveal whether there are also smaller contributions from excitations to higher shells. The cross-section rise at energies above  1860~eV is caused by direct $1s$ ionization. The theoretically predicted threshold energy for this process is 1852~eV \cite{Verner1993a} which is in reasonable accord with the experimental finding considering the fact that a photon-energy calibration has still to be applied to the experimental energy scale. It should be pointed out that PIPE is currently the only photon-ion merged-beams setup where photon energies beyond 1000~eV available.

\section{Conclusions and Outlook}

It is fair to state that the implementation of the photon-ion merged-beams technique at a world-leading synchrotron beamline has led to a breakthrough for experimental inner-shell ionization studies with ions. The high sensitivity of the PIPE setup allows one to measure cross sections on a sub-kilobarn level and, thus, to gain insight in multiple ionization processes in unprecedented detail. This facilitates the observation of rare  multi-electron processes, such as, e.g., the triple-Auger process. At present, the merged-beams method is the only experimental approach that can independently provide \emph{absolute} cross sections, which is of particular value for the application of atomic cross-section data in astrophysics. A draw-back is that the ion beams are often mixtures of ground-level and excited metastable ions and that there is no general method to quantitatively determine the ion-beam composition. In special cases, these can be inferred, e.g.,  by applying the principle of detailed balance to comparisons with experimental cross sections for photorecombination, which is the time inverse process of photoionization. In general, the use of state-prepared ion beams is desirable. Photoionization experiments with state-prepared ions in traps have been demonstrated, but the associated particle losses seem discouraging. Moreover, trap-based methods cannot generally provide absolute cross sections. Another route towards state preparation  is taken at ion storage rings, where photon-ion merged-beams experiments with portable XUV light sources based on high-harmonic generation are currently being devised. In the foreseeable future, synchrotron based  photon-ion merged-beams setups will continue to deliver the bulk of experimental data on the photoionization of ions \cite{Ueda2019}.

\vspace{6pt}

\acknowledgments{ The authors are grateful to their collaborators for their contributions  to the  work at PIPE reviewed here, who are, in alphabetical order, Sadia Bari, Dietrich Bernhardt, Alexander Borovik, Ticia Buhr, Jonas Hellhund, Pierre-Michel Hillenbrand, Kristof Holste, David Kilcoyne, Stephan Klumpp, Michael Martins, Alexander Perry-Sassmannshausen, Ron Phaneuf, Simon Reinwardt, Sandor Ricz, Daniel Savin, Jörn Seltmann, Kaja Schubert, Florian Trinter, Jens Viefhaus, and Patrick Wilhelm. We also thank Randolf Beerwerth, Stephan Fritzsche, Paul Indelicato, Eva Lindroth, and Sebastian Stock for their theory contributions to our joint work, and the staff of the PETRA\,III beamline P04 for their excellent support. Financial support was provided by the German Federal Ministry of Education and Research (BMBF,  grant numbers 05K10RG1, 05K16RG1, 05K19RG3) and by Deutsche Forschungsgemeinschaft (DFG, grant numbers Mu1068/22-1, Schi378/12-1).}


\reftitle{References}

\begin{thebibliography}{-------}
\providecommand{\natexlab}[1]{#1}

\bibitem[Pain \em{et~al.}(2017)Pain, Gilleron, and Comet]{Pain2017}
Pain, J.C.; Gilleron, F.; Comet, M.
\newblock Detailed opacity calculations for astrophysical applications.
\newblock {\em Atoms} {\bf 2017}, {\em 5},~22.
\newblock
  doi:{\changeurlcolor{black}\href{https://doi.org/10.3390/atoms5020022}{\detokenize{10.3390/atoms5020022}}}.

\bibitem[Tanaka \em{et~al.}(2020)Tanaka, Kato, Gaigalas, and
  Kawaguchi]{Tanaka2020}
Tanaka, M.; Kato, D.; Gaigalas, G.; Kawaguchi, K.
\newblock Systematic Opacity Calculations for Kilonovae.
\newblock {\em Mon. Not. R. Astron. Soc.} {\bf 2020}, {\em 496},~1369.
\newblock
  doi:{\changeurlcolor{black}\href{https://doi.org/10.1093/mnras/staa1576}{\detokenize{10.1093/mnras/staa1576}}}.

\bibitem[Paerels and Kahn(2003)]{Paerels2003a}
Paerels, F.B.S.; Kahn, S.M.
\newblock High-resolution {X}-ray spectroscopy with {C}handra and {XMM-N}ewton.
\newblock {\em Annu. Rev. Astron. Astrophys.} {\bf 2003}, {\em 41},~291--342.
\newblock
  doi:{\changeurlcolor{black}\href{https://doi.org/10.1146/annurev.astro.41.071601.165952}{\detokenize{10.1146/annurev.astro.41.071601.165952}}}.

\bibitem[Asplund \em{et~al.}(2009)Asplund, Grevesse, Sauval, and
  Scott]{Asplund2009}
Asplund, M.; Grevesse, N.; Sauval, A.J.; Scott, P.
\newblock The chemical composition of the {S}un.
\newblock {\em Annu. Rev. Astron. Astrophys.} {\bf 2009}, {\em 47},~481--522.
\newblock
  doi:{\changeurlcolor{black}\href{https://doi.org/10.1146/annurev.astro.46.060407.145222}{\detokenize{10.1146/annurev.astro.46.060407.145222}}}.

\bibitem[Ferland(2003)]{Ferland2003a}
Ferland, G.J.
\newblock Quantitative spectroscopy of astronomical plasmas.
\newblock {\em Annu. Rev. Astron. Astrophys.} {\bf 2003}, {\em 41},~517--554.

\bibitem[Kallman and Palmeri(2007)]{Kallman2007a}
Kallman, T.R.; Palmeri, P.
\newblock Atomic data for x-ray astrophysics.
\newblock {\em Rev. Mod. Phys.} {\bf 2007}, {\em 79},~79--133.
\newblock
  doi:{\changeurlcolor{black}\href{https://doi.org/10.1103/RevModPhys.79.79}{\detokenize{10.1103/RevModPhys.79.79}}}.

\bibitem[Savin \em{et~al.}(2012)Savin, Brickhouse, Cowan, Drake, Federman,
  Ferland, Frank, Gudipati, Haxton, Herbst, Profumo, Salama, Ziurys, and
  Zweibel]{Savin2012}
Savin, D.W.; Brickhouse, N.S.; Cowan, J.J.; Drake, R.P.; Federman, S.R.;
  Ferland, G.J.; Frank, A.; Gudipati, M.S.; Haxton, W.C.; Herbst, E.; Profumo,
  S.; Salama, F.; Ziurys, L.M.; Zweibel, E.G.
\newblock The impact of recent advances in laboratory astrophysics on our
  understanding of the {C}osmos.
\newblock {\em Rep. Prog. Phys.} {\bf 2012}, {\em 75},~036901.
\newblock
  doi:{\changeurlcolor{black}\href{https://doi.org/10.1088/0034-4885/75/3/036901}{\detokenize{10.1088/0034-4885/75/3/036901}}}.

\bibitem[Smith and Brickhouse(2014)]{Smith2014}
Smith, R.K.; Brickhouse, N.S.
\newblock Atomic Data Needs for Understanding X-ray Astrophysical Plasmas.
\newblock {\em Adv. At. Mol. Opt. Phys.} {\bf 2014}, {\em 63},~271 -- 321.
\newblock
  doi:{\changeurlcolor{black}\href{https://doi.org/10.1016/B978-0-12-800129-5.00004-3}{\detokenize{10.1016/B978-0-12-800129-5.00004-3}}}.

\bibitem[Lynas-Gray \em{et~al.}(2018)Lynas-Gray, Basu, Bautista, Colgan,
  Mendoza, Tennyson, Trampedach, and Turck-Chi{\`{e}}ze]{LynasGray2018}
Lynas-Gray, A.E.; Basu, S.; Bautista, M.A.; Colgan, J.; Mendoza, C.; Tennyson,
  J.; Trampedach, R.; Turck-Chi{\`{e}}ze, S.
\newblock Current State of Astrophysical Opacities: A White Paper.
\newblock {\em ASP Conf. Ser.} {\bf 2018}, {\em 515},~115.

\bibitem[Smith \em{et~al.}(2020)Smith, Hahn, Raymond, Kallman, Ballance,
  Polito, Zanna, Gu, Hell, Cumbee, Betancourt-Martinez, Costantini, and
  Corrales]{Smith2020}
Smith, R.; Hahn, M.; Raymond, J.; Kallman, T.; Ballance, C.P.; Polito, V.;
  Zanna, G.D.; Gu, L.; Hell, N.; Cumbee, R.; Betancourt-Martinez, G.;
  Costantini, E.; Corrales, L.
\newblock Roadmap on cosmic {EUV} and x-ray spectroscopy.
\newblock {\em J. Phys. B} {\bf 2020}, {\em 53},~092001.
\newblock
  doi:{\changeurlcolor{black}\href{https://doi.org/10.1088/1361-6455/ab69aa}{\detokenize{10.1088/1361-6455/ab69aa}}}.

\bibitem[Kramida \em{et~al.}(2019)Kramida, Ralchenko, Reader, and
  Team]{Kramida2019}
Kramida, A.; Ralchenko, Y.; Reader, J.; Team, N.A.
\newblock {NIST} atomic spectra database (version 5.7.1), [online]. available:
  http://physics.nist.gov/asd.
\newblock Technical report, National Institute of Standards and Technology,
  2019.
\newblock
  doi:{\changeurlcolor{black}\href{https://doi.org/10.18434/T4W30F}{\detokenize{10.18434/T4W30F}}}.

\bibitem[Kennedy \em{et~al.}(2004)Kennedy, Costello, Mosnier, and van
  Kampen]{Kennedy2004a}
Kennedy, E.T.; Costello, J.T.; Mosnier, J.P.; van Kampen, P.
\newblock {VUV/EUV} ionising radiation and atoms and ions: dual laser plasma
  investigations.
\newblock {\em Rad. Phys. Chem.} {\bf 2004}, {\em 70},~291--321.
\newblock
  doi:{\changeurlcolor{black}\href{https://doi.org/10.1016/j.radphyschem.2003.12.018}{\detokenize{10.1016/j.radphyschem.2003.12.018}}}.

\bibitem[Kravis \em{et~al.}(1991)Kravis, Church, Johnson, Meron, Jones, Levin,
  Sellin, Azuma, Mansour, Berry, and Druetta]{Kravis1991}
Kravis, S.D.; Church, D.A.; Johnson, B.M.; Meron, M.; Jones, K.W.; Levin, J.;
  Sellin, I.A.; Azuma, Y.; Mansour, N.B.; Berry, H.G.; Druetta, M.
\newblock Inner-shell photoionization of stored positive ions using synchrotron
  radiation.
\newblock {\em Phys. Rev. Lett.} {\bf 1991}, {\em 66},~2956--2959.
\newblock
  doi:{\changeurlcolor{black}\href{https://doi.org/10.1103/PhysRevLett.66.2956}{\detokenize{10.1103/PhysRevLett.66.2956}}}.

\bibitem[Thissen \em{et~al.}(2008)Thissen, Bizau, Blancard, Coreno, Dehon,
  Franceschi, Giuliani, Lemaire, and Nicolas]{Thissen2008a}
Thissen, R.; Bizau, J.M.; Blancard, C.; Coreno, M.; Dehon, C.; Franceschi, P.;
  Giuliani, A.; Lemaire, J.; Nicolas, C.
\newblock Photoionization cross section of {Xe$^+$} ion in the pure
  {$5p^5\;{^2}P_{3/2}$} ground level.
\newblock {\em Phys. Rev. Lett.} {\bf 2008}, {\em 100},~223001.
\newblock
  doi:{\changeurlcolor{black}\href{https://doi.org/10.1103/PhysRevLett.100.223001}{\detokenize{10.1103/PhysRevLett.100.223001}}}.

\bibitem[Bizau \em{et~al.}(2011)Bizau, Blancard, Coreno, Cubaynes, Dehon,
  Hassan, Folkmann, Gharaibeh, Giuliani, Lemaire, Milosavljevi\v{c}, Nicolas,
  and Thissen]{Bizau2011}
Bizau, J.M.; Blancard, C.; Coreno, M.; Cubaynes, D.; Dehon, C.; Hassan, N.E.;
  Folkmann, F.; Gharaibeh, M.F.; Giuliani, A.; Lemaire, J.; Milosavljevi\v{c},
  A.R.; Nicolas, C.; Thissen, R.
\newblock Photoionization study of {Kr$^+$} and {Xe$^+$} ions with the combined
  use of a merged-beam set-up and an ion trap.
\newblock {\em J. Phys. B} {\bf 2011}, {\em 44},~055205.

\bibitem[Simon \em{et~al.}(2010{\natexlab{a}})Simon, {Crespo
  L\'{o}pez-Urrutia}, Beilmann, Schwarz, Harman, Epp, Schmitt, Baumann, Behar,
  Bernitt, Follath, Ginzel, Keitel, Klawitter, Kubi\v{c}ek, M\"{a}ckel, Mokler,
  Reichardt, Schwarzkopf, and Ullrich]{Simon2010a}
Simon, M.C.; {Crespo L\'{o}pez-Urrutia}, J.R.; Beilmann, C.; Schwarz, M.;
  Harman, Z.; Epp, S.W.; Schmitt, B.L.; Baumann, T.M.; Behar, E.; Bernitt, S.;
  Follath, R.; Ginzel, R.; Keitel, C.H.; Klawitter, R.; Kubi\v{c}ek, K.;
  M\"{a}ckel, V.; Mokler, P.H.; Reichardt, G.; Schwarzkopf, O.; Ullrich, J.
\newblock Resonant and near-threshold photoionization cross sections of
  {Fe$^{14+}$}.
\newblock {\em Phys. Rev. Lett.} {\bf 2010}, {\em 105},~183001.
\newblock
  doi:{\changeurlcolor{black}\href{https://doi.org/10.1103/PhysRevLett.105.183001}{\detokenize{10.1103/PhysRevLett.105.183001}}}.

\bibitem[Simon \em{et~al.}(2010{\natexlab{b}})Simon, Schwarz, Epp, Beilmann,
  Schmitt, Harman, Baumann, Mokler, Bernitt, Ginzel, Higgins, Keitel,
  Klawitter, Kubi{\v{c}}ek, M\"{a}ckel, Ullrich, and Crespo L{\'
  o}pez-Urrutia]{Simon2010}
Simon, M.C.; Schwarz, M.; Epp, S.W.; Beilmann, C.; Schmitt, B.L.; Harman, Z.;
  Baumann, T.M.; Mokler, P.H.; Bernitt, S.; Ginzel, R.; Higgins, S.G.; Keitel,
  C.H.; Klawitter, R.; Kubi{\v{c}}ek, K.; M\"{a}ckel, V.; Ullrich, J.; Crespo
  L{\' o}pez-Urrutia, J.R.
\newblock Photoionization of {N$^{3 +}$} and {Ar$^{8 +}$} in an electron beam
  ion trap by synchrotron radiation.
\newblock {\em J. Phys. B} {\bf 2010}, {\em 43},~065003.
\newblock
  doi:{\changeurlcolor{black}\href{https://doi.org/10.1088/0953-4075/43/6/065003}{\detokenize{10.1088/0953-4075/43/6/065003}}}.

\bibitem[Hirsch \em{et~al.}(2012)Hirsch, Zamudio-Bayer, Ameseder, Langenberg,
  Rittmann, Vogel, M\"{o}ller, v.~Issendorff, and Lau]{Hirsch2012}
Hirsch, K.; Zamudio-Bayer, V.; Ameseder, F.; Langenberg, A.; Rittmann, J.;
  Vogel, M.; M\"{o}ller, T.; v.~Issendorff, B.; Lau, J.T.
\newblock 2p x-ray absorption of free transition-metal cations across the 3d
  transition elements: calcium through copper.
\newblock {\em Phys. Rev. A} {\bf 2012}, {\em 85},~062501.
\newblock
  doi:{\changeurlcolor{black}\href{https://doi.org/10.1103/PhysRevA.85.062501}{\detokenize{10.1103/PhysRevA.85.062501}}}.

\bibitem[Lyon \em{et~al.}(1986)Lyon, Peart, West, and Dolder]{Lyon1986}
Lyon, I.C.; Peart, B.; West, J.B.; Dolder, K.
\newblock Measurements of absolute cross sections for the photoionisation of
  {Ba$^{+}$} ions.
\newblock {\em J. Phys. B} {\bf 1986}, {\em 19},~4137--4147.

\bibitem[Kjeldsen(2006)]{Kjeldsen2006a}
Kjeldsen, H.
\newblock Photoionization cross sections of atomic ions from merged-beam
  experiments.
\newblock {\em J. Phys. B} {\bf 2006}, {\em 39},~R325--R377.
\newblock
  doi:{\changeurlcolor{black}\href{https://doi.org/10.1088/0953-4075/39/21/R01}{\detokenize{10.1088/0953-4075/39/21/R01}}}.

\bibitem[Schippers \em{et~al.}(2016)Schippers, Kilcoyne, Phaneuf, and
  M\"{u}ller]{Schippers2016}
Schippers, S.; Kilcoyne, A.L.D.; Phaneuf, R.A.; M\"{u}ller, A.
\newblock Photoionization of ions with synchrotron radiation: from ions in
  space to atoms in cages.
\newblock {\em Contemp. Phys.} {\bf 2016}, {\em 57},~215--229.
\newblock
  doi:{\changeurlcolor{black}\href{https://doi.org/10.1080/00107514.2015.1109771}{\detokenize{10.1080/00107514.2015.1109771}}}.

\bibitem[Bizau \em{et~al.}(1991)Bizau, Cubaynes, Richter, Wuilleumier, Obert,
  Putaux, Morgan, K\"{a}llne, Sorensen, and Damany]{Bizau1991}
Bizau, J.M.; Cubaynes, D.; Richter, M.; Wuilleumier, F.J.; Obert, J.; Putaux,
  J.C.; Morgan, T.J.; K\"{a}llne, E.; Sorensen, S.; Damany, A.
\newblock First observation of photoelectron spectra emitted in the
  photoionization of a singly charged-ion beam with synchrotron radiation.
\newblock {\em Phys. Rev. Lett.} {\bf 1991}, {\em 67},~576--579.

\bibitem[Oura \em{et~al.}(1994)Oura, Kravis, Koizumi, Itoh, Kojima, Sano,
  Sekioka, Kimura, Okuno, and Awaya]{Oura1994}
Oura, M.; Kravis, S.; Koizumi, T.; Itoh, Y.; Kojima, T.M.; Sano, M.; Sekioka,
  T.; Kimura, M.; Okuno, K.; Awaya, Y.
\newblock Experimental setups for photoionization of multiply charged ions by
  synchrotron radiation.
\newblock {\em Nucl. Instrum. Methods B} {\bf 1994}, {\em 86},~190.
\newblock
  doi:{\changeurlcolor{black}\href{https://doi.org/10.1016/0168-583X(94)96174-3}{\detokenize{10.1016/0168-583X(94)96174-3}}}.

\bibitem[Kjeldsen \em{et~al.}(1999)Kjeldsen, Folkmann, Knudsen, Rasmussen,
  West, and Andersen]{Kjeldsen1999b}
Kjeldsen, H.; Folkmann, F.; Knudsen, H.; Rasmussen, M.S.; West, J.B.; Andersen,
  T.
\newblock Absolute photoionization cross cection of {K$^{+}$} ions from the
  {$3p$} to the {$3s$} threshold.
\newblock {\em J. Phys. B} {\bf 1999}, {\em 32},~4457--4465.
\newblock
  doi:{\changeurlcolor{black}\href{https://doi.org/10.1088/0953-4075/32/18/306}{\detokenize{10.1088/0953-4075/32/18/306}}}.

\bibitem[Yamaoka \em{et~al.}(2001)Yamaoka, Oura, Kawatsura, Hayaishi, Sekioka,
  Agui, Yoshigoe, and Koike]{Yamaoka2001}
Yamaoka, H.; Oura, M.; Kawatsura, K.; Hayaishi, T.; Sekioka, T.; Agui, A.;
  Yoshigoe, A.; Koike, F.
\newblock Photoionization of singly and doubly charged neon ions following
  inner-shell excitation.
\newblock {\em Phys. Rev. A} {\bf 2001}, {\em 65},~012709.
\newblock
  doi:{\changeurlcolor{black}\href{https://doi.org/10.1103/PhysRevA.65.012709}{\detokenize{10.1103/PhysRevA.65.012709}}}.

\bibitem[Covington \em{et~al.}(2002)Covington, Aguilar, Covington, Gharaibeh,
  Hinojosa, Shirley, Phaneuf, \'{A}lvarez, Cisneros, Dominguez-Lopez,
  Sant{'}Anna, Schlachter, McLaughlin, and Dalgarno]{Covington2002}
Covington, A.M.; Aguilar, A.; Covington, I.R.; Gharaibeh, M.F.; Hinojosa, G.;
  Shirley, C.A.; Phaneuf, R.A.; \'{A}lvarez, I.; Cisneros, C.; Dominguez-Lopez,
  I.; Sant{'}Anna, M.M.; Schlachter, A.S.; McLaughlin, B.M.; Dalgarno, A.
\newblock Photoionization of {Ne$^+$} using synchrotron radiation.
\newblock {\em Phys. Rev. A} {\bf 2002}, {\em 66},~062710.

\bibitem[Schippers \em{et~al.}(2014)Schippers, Ricz, Buhr, Borovik, Hellhund,
  Holste, Huber, Sch\"{a}fer, Schury, Klumpp, Mertens, Martins, Flesch, Ulrich,
  R\"{u}hl, Jahnke, Lower, Metz, Schmidt, Sch\"{o}ffler, Williams, Glaser,
  Scholz, Seltmann, Viefhaus, Dorn, Wolf, Ullrich, and
  M\"{u}ller]{Schippers2014}
Schippers, S.; Ricz, S.; Buhr, T.; Borovik, Jr., A.; Hellhund, J.; Holste, K.;
  Huber, K.; Sch\"{a}fer, H.J.; Schury, D.; Klumpp, S.; Mertens, K.; Martins,
  M.; Flesch, R.; Ulrich, G.; R\"{u}hl, E.; Jahnke, T.; Lower, J.; Metz, D.;
  Schmidt, L.P.H.; Sch\"{o}ffler, M.; Williams, J.B.; Glaser, L.; Scholz, F.;
  Seltmann, J.; Viefhaus, J.; Dorn, A.; Wolf, A.; Ullrich, J.; M\"{u}ller, A.
\newblock Absolute cross sections for photoionization of {Xe$^{q+}$} ions ({$1
  \le q \le 5$}) at the 3d ionization threshold.
\newblock {\em J. Phys. B} {\bf 2014}, {\em 47},~115602.
\newblock
  doi:{\changeurlcolor{black}\href{https://doi.org/10.1088/0953-4075/47/11/115602}{\detokenize{10.1088/0953-4075/47/11/115602}}}.

\bibitem[Bizau \em{et~al.}(2016)Bizau, Cubaynes, Guilbaud, Eassan, Shorman,
  Bouisset, Guigand, Moustier, Mari\'{e}, Nadal, Robert, Nicolas, and
  Miron]{Bizau2016a}
Bizau, J.M.; Cubaynes, D.; Guilbaud, S.; Eassan, N.E.; Shorman, M.M.A.;
  Bouisset, E.; Guigand, J.; Moustier, O.; Mari\'{e}, A.; Nadal, E.; Robert,
  E.; Nicolas, C.; Miron, C.
\newblock A merged-beam setup at {SOLEIL} dedicated to photoelectron-photoion
  coincidence studies on ionic species.
\newblock {\em J. Elec. Spectrosc. Rel. Phenom.} {\bf 2016}, {\em 210},~5--12.
\newblock
  doi:{\changeurlcolor{black}\href{https://doi.org/10.1016/j.elspec.2016.03.006}{\detokenize{10.1016/j.elspec.2016.03.006}}}.

\bibitem[Lestinsky \em{et~al.}(2016)Lestinsky, Andrianov, Aurand, Bagnoud,
  Bernhardt, Beyer, Bishop, Blaum, Bleile, Borovik, Bosch, Bostock, Brandau,
  Br\"{a}uning-Demian, Bray, Davinson, Ebinger, Echler, Egelhof, Ehresmann,
  Engstr\"{o}m, Enss, Ferreira, Fischer, Fleischmann, F\"{o}rster, Fritzsche,
  Geithner, Geyer, Glorius, G\"{o}bel, Gorda, Goullon, Grabitz, Grisenti,
  Gumberidze, Hagmann, Heil, Heinz, Herfurth, He{\ss}, Hillenbrand, Hubele,
  Indelicato, K\"{a}llberg, Kester, Kiselev, Knie, Kozhuharov, Kraft-Bermuth,
  K\"{u}hl, Lane, Litvinov, Liesen, Ma, M\"{a}rtin, Moshammer, M\"{u}ller,
  Namba, Neumeyer, Nilsson, N\"{o}rtersh\"{a}user, Paulus, Petridis, Reed,
  Reifarth, Rei{\ss}, Rothhardt, Sanchez, Sanjari, Schippers, Schmidt,
  Schneider, Scholz, Schuch, Schulz, Shabaev, Simonsson, Sj\"{o}holm,
  Skeppstedt, Sonnabend, Spillmann, Stiebing, Steck, St\"{o}hlker, Surzhykov,
  Torilov, Tr\"{a}bert, Trassinelli, Trotsenko, Tu, Uschmann, Walker, Weber,
  Winters, Woods, Zhao, and Zhang]{Lestinsky2016}
Lestinsky, M.; Andrianov, V.; Aurand, B.; Bagnoud, V.; Bernhardt, D.; Beyer,
  H.; Bishop, S.; Blaum, K.; Bleile, A.; Borovik, A.; Bosch, F.; Bostock, C.J.;
  Brandau, C.; Br\"{a}uning-Demian, A.; Bray, I.; Davinson, T.; Ebinger, B.;
  Echler, A.; Egelhof, P.; Ehresmann, A.; Engstr\"{o}m, M.; Enss, C.; Ferreira,
  N.; Fischer, D.; Fleischmann, A.; F\"{o}rster, E.; Fritzsche, S.; Geithner,
  R.; Geyer, S.; Glorius, J.; G\"{o}bel, K.; Gorda, O.; Goullon, J.; Grabitz,
  P.; Grisenti, R.; Gumberidze, A.; Hagmann, S.; Heil, M.; Heinz, A.; Herfurth,
  F.; He{\ss}, R.; Hillenbrand, P.M.; Hubele, R.; Indelicato, P.; K\"{a}llberg,
  A.; Kester, O.; Kiselev, O.; Knie, A.; Kozhuharov, C.; Kraft-Bermuth, S.;
  K\"{u}hl, T.; Lane, G.; Litvinov, Y.A.; Liesen, D.; Ma, X.W.; M\"{a}rtin, R.;
  Moshammer, R.; M\"{u}ller, A.; Namba, S.; Neumeyer, P.; Nilsson, T.;
  N\"{o}rtersh\"{a}user, W.; Paulus, G.; Petridis, N.; Reed, M.; Reifarth, R.;
  Rei{\ss}, P.; Rothhardt, J.; Sanchez, R.; Sanjari, M.S.; Schippers, S.;
  Schmidt, H.T.; Schneider, D.; Scholz, P.; Schuch, R.; Schulz, M.; Shabaev,
  V.; Simonsson, A.; Sj\"{o}holm, J.; Skeppstedt, O.; Sonnabend, K.; Spillmann,
  U.; Stiebing, K.; Steck, M.; St\"{o}hlker, T.; Surzhykov, A.; Torilov, S.;
  Tr\"{a}bert, E.; Trassinelli, M.; Trotsenko, S.; Tu, X.L.; Uschmann, I.;
  Walker, P.M.; Weber, G.; Winters, D.F.A.; Woods, P.J.; Zhao, H.Y.; Zhang,
  Y.H.
\newblock Physics book: {CRYRING@ESR}.
\newblock {\em Eur. Phys. J. ST} {\bf 2016}, {\em 225},~797--882.
\newblock
  doi:{\changeurlcolor{black}\href{https://doi.org/10.1140/epjst/e2016-02643-6}{\detokenize{10.1140/epjst/e2016-02643-6}}}.

\bibitem[Borovik \em{et~al.}(2020)Borovik, Weber, Hilbert, Lin,
  Pf{\"{a}}fflein, Zhu, Hahn, Lestinsky, Schippers, St{\"{o}}hlker, and
  Rothhardt]{Borovik2020}
Borovik, A.; Weber, G.; Hilbert, V.; Lin, H.; Pf{\"{a}}fflein, P.; Zhu, B.;
  Hahn, C.; Lestinsky, M.; Schippers, S.; St{\"{o}}hlker, T.; Rothhardt, J.
\newblock Development of a detector to register low-energy, charge-changed ions
  from ionization experiments at {CRYRING}@{ESR}.
\newblock {\em J. Phys.: Conf. Ser.} {\bf 2020}, {\em 1412},~242003.
\newblock
  doi:{\changeurlcolor{black}\href{https://doi.org/10.1088/1742-6596/1412/24/242003}{\detokenize{10.1088/1742-6596/1412/24/242003}}}.

\bibitem[Viefhaus \em{et~al.}(2013)Viefhaus, Scholz, Deinert, Glaser, Ilchen,
  Seltmann, Walter, and Siewert]{Viefhaus2013}
Viefhaus, J.; Scholz, F.; Deinert, S.; Glaser, L.; Ilchen, M.; Seltmann, J.;
  Walter, P.; Siewert, F.
\newblock The variable polarization {XUV} beamline {P04} at {PETRA} {III}:
  optics, mechanics and their performance.
\newblock {\em Nucl. Instrum. Methods A} {\bf 2013}, {\em 710},~151--154.
\newblock
  doi:{\changeurlcolor{black}\href{https://doi.org/10.1016/j.nima.2012.10.110}{\detokenize{10.1016/j.nima.2012.10.110}}}.

\bibitem[Schippers \em{et~al.}(2020)Schippers, Buhr, Borovik, Holste,
  Perry-Sassmannshausen, Mertens, Reinwardt, Martins, Klumpp, Schubert, Bari,
  Beerwerth, Fritzsche, Ricz, Hellhund, and M\"{u}ller]{Schippers2020}
Schippers, S.; Buhr, T.; Borovik, Jr., A.; Holste, K.; Perry-Sassmannshausen,
  A.; Mertens, K.; Reinwardt, S.; Martins, M.; Klumpp, S.; Schubert, K.; Bari,
  S.; Beerwerth, R.; Fritzsche, S.; Ricz, S.; Hellhund, J.; M\"{u}ller, A.
\newblock The photon-ion merged-beams experiment {PIPE} at {PETRA\,III} - {T}he
  first five years.
\newblock {\em X-Ray Spectrometry} {\bf 2020}, {\em 49},~11.
\newblock
  doi:{\changeurlcolor{black}\href{https://doi.org/10.1002/xrs.3035}{\detokenize{10.1002/xrs.3035}}}.

\bibitem[M\"{u}ller \em{et~al.}(2017)M\"{u}ller, Bernhardt, Borovik, Buhr,
  Hellhund, Holste, Kilcoyne, Klumpp, Martins, Ricz, Seltmann, Viefhaus, and
  Schippers]{Mueller2017}
M\"{u}ller, A.; Bernhardt, D.; Borovik, Jr., A.; Buhr, T.; Hellhund, J.;
  Holste, K.; Kilcoyne, A.L.D.; Klumpp, S.; Martins, M.; Ricz, S.; Seltmann,
  J.; Viefhaus, J.; Schippers, S.
\newblock Photoionization of {N}e atoms and {Ne$^{+}$} ions near the {K} edge:
  precision spectroscopy and absolute cross sections.
\newblock {\em Astrophys. J.} {\bf 2017}, {\em 836},~166.
\newblock
  doi:{\changeurlcolor{black}\href{https://doi.org/10.3847/1538-4357/836/2/166}{\detokenize{10.3847/1538-4357/836/2/166}}}.

\bibitem[M\"{u}ller \em{et~al.}(2018)M\"{u}ller, Lindroth, Bari, Borovik~Jr.,
  Hillenbrand, Holste, Indelicato, Kilcoyne, Klumpp, Martins, Viefhaus,
  Wilhelm, and Schippers]{Mueller2018c}
M\"{u}ller, A.; Lindroth, E.; Bari, S.; Borovik~Jr., A.; Hillenbrand, P.M.;
  Holste, K.; Indelicato, P.; Kilcoyne, A.L.D.; Klumpp, S.; Martins, M.;
  Viefhaus, J.; Wilhelm, P.; Schippers, S.
\newblock Photoionization of metastable heliumlike {C$^{4+}$($1s\,2s\;^3S_1$)}
  ions: precision study of intermediate doubly excited states.
\newblock {\em Phys. Rev. A} {\bf 2018}, {\em 98},~033416.
\newblock
  doi:{\changeurlcolor{black}\href{https://doi.org/10.1103/PhysRevA.98.033416}{\detokenize{10.1103/PhysRevA.98.033416}}}.

\bibitem[Yerokhin \em{et~al.}(2017{\natexlab{a}})Yerokhin, Surzhykov, and
  M\"{u}ller]{Yerokhin2017}
Yerokhin, V.A.; Surzhykov, A.; M\"{u}ller, A.
\newblock Relativistic configuration-interaction calculations of the energy
  levels of the {1s$^{2}$2l} and 1s 2l 2l' states in lithiumlike ions: carbon
  through chlorine.
\newblock {\em Phys. Rev. A} {\bf 2017}, {\em 96},~042505.
\newblock
  doi:{\changeurlcolor{black}\href{https://doi.org/10.1103/PhysRevA.96.042505}{\detokenize{10.1103/PhysRevA.96.042505}}}.

\bibitem[Yerokhin \em{et~al.}(2017{\natexlab{b}})Yerokhin, Surzhykov, and
  M\"{u}ller]{Yerokhin2017a}
Yerokhin, V.A.; Surzhykov, A.; M\"{u}ller, A.
\newblock Erratum: relativistic configuration-interaction calculations of the
  energy levels of the {1s$^{2}$2l} and 1s 2l 2l' states in lithiumlike ions:
  carbon through chlorine {[Phys. Rev. A 96, 042505 (2017)]}.
\newblock {\em Phys. Rev. A} {\bf 2017}, {\em 96},~069901.
\newblock
  doi:{\changeurlcolor{black}\href{https://doi.org/10.1103/PhysRevA.96.069901}{\detokenize{10.1103/PhysRevA.96.069901}}}.

\bibitem[Machado \em{et~al.}(2020)Machado, Bian, Paul, Trassinelli, Amaro,
  Guerra, Szabo, Gumberidze, Isac, Santos, Desclaux, and
  Indelicato]{Machado2020}
Machado, J.; Bian, G.; Paul, N.; Trassinelli, M.; Amaro, P.; Guerra, M.; Szabo,
  C.I.; Gumberidze, A.; Isac, J.M.; Santos, J.P.; Desclaux, J.P.; Indelicato,
  P.
\newblock Reference-free measurements of the
  {$1s^2\,2s\,2p\;^{2}P_{1/2,3/2}^{o} \to1{s}^{2}\,2s\;^{2}{S}_{1/2}$} and
  {$1s\,2s\,2p\;^{4}{P}_{5/2} \to 1{s}^{2}\,2s\;^{2}{S}_{1/2}$} transition
  energies and widths in lithiumlike sulfur and argon ions.
\newblock {\em Phys. Rev. A} {\bf 2020}, {\em 101},~062505.
\newblock
  doi:{\changeurlcolor{black}\href{https://doi.org/10.1103/PhysRevA.101.062505}{\detokenize{10.1103/PhysRevA.101.062505}}}.

\bibitem[Schippers \em{et~al.}(2017)Schippers, Martins, Beerwerth, Bari,
  Holste, Schubert, Viefhaus, Savin, Fritzsche, and M\"{u}ller]{Schippers2017}
Schippers, S.; Martins, M.; Beerwerth, R.; Bari, S.; Holste, K.; Schubert, K.;
  Viefhaus, J.; Savin, D.W.; Fritzsche, S.; M\"{u}ller, A.
\newblock Near {L}-edge single and multiple photoionization of singly charged
  iron ions.
\newblock {\em Astrophys. J.} {\bf 2017}, {\em 849},~5.
\newblock
  doi:{\changeurlcolor{black}\href{https://doi.org/10.3847/1538-4357/aa8fcc}{\detokenize{10.3847/1538-4357/aa8fcc}}}.

\bibitem[Schippers \em{et~al.}()Schippers, Beerwerth, Bari, Buhr, Holste,
  Kilcoyne, Perry-Sassmannshausen, Phaneuf, Reinwardt, Savin, Schubert,
  Fritzsche, Martins, and M{\"{u}}ller]{Schippers2020b}
Schippers, S.; Beerwerth, R.; Bari, S.; Buhr, T.; Holste, K.; Kilcoyne, A.L.D.;
  Perry-Sassmannshausen, A.; Phaneuf, R.A.; Reinwardt, S.; Savin, D.W.;
  Schubert, K.; Fritzsche, S.; Martins, M.; M{\"{u}}ller, A.
\newblock Near {L}-edge single and multiple photoionization of doubly charged
  iron ions.
\newblock in preparation.

\bibitem[Beerwerth \em{et~al.}(2019)Beerwerth, Buhr, Perry-Sassmannshausen,
  Stock, Bari, Holste, Kilcoyne, Reinwardt, Ricz, Savin, Schubert, Martins,
  Müller, Fritzsche, and Schippers]{Beerwerth2019}
Beerwerth, R.; Buhr, T.; Perry-Sassmannshausen, A.; Stock, S.O.; Bari, S.;
  Holste, K.; Kilcoyne, A.L.D.; Reinwardt, S.; Ricz, S.; Savin, D.W.; Schubert,
  K.; Martins, M.; Müller, A.; Fritzsche, S.; Schippers, S.
\newblock Near {L}-edge single and multiple photoionization of triply charged
  iron ions.
\newblock {\em Astrophys. J.} {\bf 2019}, {\em 887},~189.
\newblock
  doi:{\changeurlcolor{black}\href{https://doi.org/10.3847/1538-4357/ab5118}{\detokenize{10.3847/1538-4357/ab5118}}}.

\bibitem[Jensen and Snow(2007)]{Jensen2007a}
Jensen, A.G.; Snow, T.P.
\newblock New insights on interstellar gas-phase iron.
\newblock {\em Astrophys. J.} {\bf 2007}, {\em 669},~378--400.
\newblock
  doi:{\changeurlcolor{black}\href{https://doi.org/10.1086/521638}{\detokenize{10.1086/521638}}}.

\bibitem[Jenkins(2009)]{Jenkins2009}
Jenkins, E.B.
\newblock A unified representation of gas-phase element depletions in the
  interstellar medium.
\newblock {\em Astrophys. J.} {\bf 2009}, {\em 700},~1299.
\newblock
  doi:{\changeurlcolor{black}\href{https://doi.org/10.1088/0004-637X/700/2/1299}{\detokenize{10.1088/0004-637X/700/2/1299}}}.

\bibitem[Juett \em{et~al.}(2006)Juett, Schulz, Chakrabarty, and
  Gorczyca]{Juett2006}
Juett, A.M.; Schulz, N.S.; Chakrabarty, D.; Gorczyca, T.W.
\newblock High-resolution x-ray spectroscopy of the interstellar medium. {II}.
  neon and iron absorption edges.
\newblock {\em Astrophys. J.} {\bf 2006}, {\em 648},~1066.
\newblock
  doi:{\changeurlcolor{black}\href{https://doi.org/10.1086/506189}{\detokenize{10.1086/506189}}}.

\bibitem[Miedema and de~Groot(2013)]{Miedema2013}
Miedema, P.S.; de~Groot, F.M.F.
\newblock The iron {L} edges:{F}e 2p x-ray absorption and electron energy loss
  spectroscopy.
\newblock {\em J. Elec. Spectrosc. Rel. Phenom.} {\bf 2013}, {\em 187},~32--48.
\newblock
  doi:{\changeurlcolor{black}\href{https://doi.org/10.1016/j.elspec.2013.03.005}{\detokenize{10.1016/j.elspec.2013.03.005}}}.

\bibitem[Richter \em{et~al.}(2004)Richter, Godehusen, Martins, Wolff, and
  Zimmermann]{Richter2004}
Richter, T.; Godehusen, K.; Martins, M.; Wolff, T.; Zimmermann, P.
\newblock Interplay of intra-atomic and interatomic effects: an investigation
  of the {$2p$} core level spectra of atomic {F}e and molecular {FeCl$_2$}.
\newblock {\em Phys. Rev. Lett.} {\bf 2004}, {\em 93},~023002.
\newblock
  doi:{\changeurlcolor{black}\href{https://doi.org/10.1103/PhysRevLett.93.023002}{\detokenize{10.1103/PhysRevLett.93.023002}}}.

\bibitem[Martins \em{et~al.}(2006)Martins, Godehusen, Richter, Wernet, and
  Zimmermann]{Martins2006a}
Martins, M.; Godehusen, K.; Richter, T.; Wernet, P.; Zimmermann, P.
\newblock Open shells and multi-electron interactions: {C}ore level
  photoionization of the 3d metal atoms.
\newblock {\em J. Phys. B} {\bf 2006}, {\em 39},~R79--R125.
\newblock
  doi:{\changeurlcolor{black}\href{https://doi.org/10.1088/0953-4075/39/5/R01}{\detokenize{10.1088/0953-4075/39/5/R01}}}.

\bibitem[Blancard \em{et~al.}(2018)Blancard, Cubaynes, Guilbaud, and
  Bizau]{Blancard2018}
Blancard, C.; Cubaynes, D.; Guilbaud, S.; Bizau, J.M.
\newblock Absolute photoionization cross section for {Fe$^{6+}$} to
  {Fe$^{10+}$} ions in the photon energy region of the 2p-3d resonance lines.
\newblock {\em Astrophys. J.} {\bf 2018}, {\em 853},~32.
\newblock
  doi:{\changeurlcolor{black}\href{https://doi.org/10.3847/1538-4357/aa9ff7}{\detokenize{10.3847/1538-4357/aa9ff7}}}.

\bibitem[Kjeldsen \em{et~al.}(2002)Kjeldsen, Kristensen, Folkmann, and
  Andersen]{Kjeldsen2002c}
Kjeldsen, H.; Kristensen, B.; Folkmann, F.; Andersen, T.
\newblock Measurements of the absolute photoionization cross section of
  {Fe$^{+}$} ions from 15.8 to 180 e{V}.
\newblock {\em J. Phys. B} {\bf 2002}, {\em 35},~3655--3668.
\newblock
  doi:{\changeurlcolor{black}\href{https://doi.org/10.1088/0953-4075/35/17/303}{\detokenize{10.1088/0953-4075/35/17/303}}}.

\bibitem[Bizau \em{et~al.}(2006)Bizau, Blancard, Cubaynes, Folkmann, Kilbane,
  Faussurier, Luna, Lemaire, Blieck, and Wuilleumier]{Bizau2006a}
Bizau, J.M.; Blancard, C.; Cubaynes, D.; Folkmann, F.; Kilbane, D.; Faussurier,
  G.; Luna, H.; Lemaire, J.L.; Blieck, J.; Wuilleumier, F.J.
\newblock Experimental and theoretical studies of the photoionization cross
  section of {Fe$^{4+}$}.
\newblock {\em Phys. Rev. A} {\bf 2006}, {\em 73},~020707.
\newblock
  doi:{\changeurlcolor{black}\href{https://doi.org/10.1103/PhysRevA.73.020707}{\detokenize{10.1103/PhysRevA.73.020707}}}.

\bibitem[El~Hassan \em{et~al.}(2009)El~Hassan, Bizau, Blancard, Cosse,
  Cubaynes, Faussurier, and Folkmann]{ElHassan2009}
El~Hassan, N.; Bizau, J.M.; Blancard, C.; Cosse, P.; Cubaynes, D.; Faussurier,
  G.; Folkmann, F.
\newblock Photoionization cross sections of iron isonuclear sequence ions:
  {Fe$^{2+}$} to {Fe$^{6+}$}.
\newblock {\em Phys. Rev. A} {\bf 2009}, {\em 79},~033415.
\newblock
  doi:{\changeurlcolor{black}\href{https://doi.org/10.1103/PhysRevA.79.033415}{\detokenize{10.1103/PhysRevA.79.033415}}}.

\bibitem[Gharaibeh \em{et~al.}(2011)Gharaibeh, Aguilar, Covington, Emmons,
  Scully, Phaneuf, M\"{u}ller, Bozek, Kilcoyne, Schlachter, \'{A}lvarez,
  Cisneros, and Hinojosa]{Gharaibeh2011}
Gharaibeh, M.F.; Aguilar, A.; Covington, A.M.; Emmons, E.D.; Scully, S.W.J.;
  Phaneuf, R.A.; M\"{u}ller, A.; Bozek, J.D.; Kilcoyne, A.L.D.; Schlachter,
  A.S.; \'{A}lvarez, I.; Cisneros, C.; Hinojosa, G.
\newblock Photoionization measurements for the iron isonuclear sequence
  {Fe$^{3+}$}, {Fe$^{5+}$}, and {Fe$^{7+}$}.
\newblock {\em Phys. Rev. A} {\bf 2011}, {\em 83},~043412.
\newblock
  doi:{\changeurlcolor{black}\href{https://doi.org/10.1103/PhysRevA.83.043412}{\detokenize{10.1103/PhysRevA.83.043412}}}.

\bibitem[Verner \em{et~al.}(1993)Verner, Yakovlev, Band, and
  Trzhaskovskaya.]{Verner1993a}
Verner, D.A.; Yakovlev, D.G.; Band, I.M.; Trzhaskovskaya., M.B.
\newblock Subshell photoionization cross sections and ionization energies of
  atoms and ions from {H}e to {Z}n.
\newblock {\em At. Data Nucl. Data Tables} {\bf 1993}, {\em 55},~233--280.
\newblock
  doi:{\changeurlcolor{black}\href{https://doi.org/10.1006/adnd.1993.1022}{\detokenize{10.1006/adnd.1993.1022}}}.

\bibitem[Kaastra and Mewe(1993)]{Kaastra1993}
Kaastra, J.S.; Mewe, R.
\newblock {X-ray emission from thin plasmas. {I} - Multiple {A}uger ionisation
  and fluorescence processes for {B}e to {Z}n}.
\newblock {\em Astron. Astrophys. Suppl. Ser.} {\bf 1993}, {\em 97},~443--482.

\bibitem[Ku{\v{c}}as \em{et~al.}(2019)Ku{\v{c}}as, Drabu{\v{z}}inskis,
  Kynien{\.{e}}, Masys, and Jonauskas]{Kucas2019}
Ku{\v{c}}as, S.; Drabu{\v{z}}inskis, P.; Kynien{\.{e}}, A.; Masys, {\v{S}}.;
  Jonauskas, V.
\newblock Evolution of radiative and Auger cascades following {$2s$} vacancy
  creation in {Fe$^{2+}$}.
\newblock {\em J. Phys. B} {\bf 2019}, {\em 52},~225001.
\newblock
  doi:{\changeurlcolor{black}\href{https://doi.org/10.1088/1361-6455/ab46fa}{\detokenize{10.1088/1361-6455/ab46fa}}}.

\bibitem[Ku{\v{c}}as \em{et~al.}(2020)Ku{\v{c}}as, Drabu{\v{z}}inskis, and
  Jonauskas]{Kucas2020}
Ku{\v{c}}as, S.; Drabu{\v{z}}inskis, P.; Jonauskas, V.
\newblock Radiative and {A}uger cascade following {$2p$} vacancy creation in
  {Fe$^{2+}$}.
\newblock {\em At. Data Nucl. Data Tables} {\bf 2020}, p. 101357.
\newblock
  doi:{\changeurlcolor{black}\href{https://doi.org/10.1016/j.adt.2020.101357}{\detokenize{10.1016/j.adt.2020.101357}}}.

\bibitem[Andersen \em{et~al.}(2001)Andersen, Andersen, Folkmann, Ivanov,
  Kjeldsen, and West]{Andersen2001b}
Andersen, P.; Andersen, T.; Folkmann, F.; Ivanov, V.K.; Kjeldsen, H.; West,
  J.B.
\newblock Absolute cross sections for the photoionization of 4d electrons in
  {Xe$^{+}$} and {Xe$^{2+}$} ions.
\newblock {\em J. Phys. B} {\bf 2001}, {\em 34},~2009--2019.
\newblock
  doi:{\changeurlcolor{black}\href{https://doi.org/10.1088/0953-4075/34/10/314}{\detokenize{10.1088/0953-4075/34/10/314}}}.

\bibitem[Berrah \em{et~al.}(2002)Berrah, Bozek, Turri, Akerman, Rude, Zhou, and
  Manson]{Berrah2002}
Berrah, N.; Bozek, J.D.; Turri, G.; Akerman, G.; Rude, B.; Zhou, H.L.; Manson,
  S.T.
\newblock K-shell photodetachment of {H}e$^-$: {E}xperiment and theory.
\newblock {\em Phys. Rev. Lett.} {\bf 2002}, {\em 88},~093001.
\newblock
  doi:{\changeurlcolor{black}\href{https://doi.org/10.1103/PhysRevLett.88.093001}{\detokenize{10.1103/PhysRevLett.88.093001}}}.

\bibitem[Bilodeau \em{et~al.}(2004)Bilodeau, Bozek, Aguilar, Ackerman, Turri,
  and Berrah]{Bilodeau2004a}
Bilodeau, R.C.; Bozek, J.D.; Aguilar, A.; Ackerman, G.D.; Turri, G.; Berrah, N.
\newblock Photoexcitation of {He$^-$} hollow-ion resonances: {O}bservation of
  the {$2s\,2p^2\,\,^4P$} state.
\newblock {\em Phys. Rev. Lett.} {\bf 2004}, {\em 93},~193001.
\newblock
  doi:{\changeurlcolor{black}\href{https://doi.org/10.1103/PhysRevLett.93.193001}{\detokenize{10.1103/PhysRevLett.93.193001}}}.

\bibitem[Kjeldsen \em{et~al.}(2001)Kjeldsen, Andersen, Folkmann, Kristensen,
  and Andersen]{Kjeldsen2001a}
Kjeldsen, H.; Andersen, P.; Folkmann, F.; Kristensen, B.; Andersen, T.
\newblock Inner-shell photodetachment of {Li$^{-}$}.
\newblock {\em J. Phys. B} {\bf 2001}, {\em 34},~L353--L357.
\newblock
  doi:{\changeurlcolor{black}\href{https://doi.org/10.1088/0953-4075/34/10/107}{\detokenize{10.1088/0953-4075/34/10/107}}}.

\bibitem[Berrah \em{et~al.}(2001)Berrah, Bozek, Wills, Turri, Zhou, Manson,
  Akerman, Rude, Gibson, Walter, VoKy, Hibbert, and Ferguson]{Berrah2001}
Berrah, N.; Bozek, J.D.; Wills, A.A.; Turri, G.; Zhou, H.L.; Manson, S.T.;
  Akerman, G.; Rude, B.; Gibson, N.D.; Walter, C.W.; VoKy, L.; Hibbert, A.;
  Ferguson, S.M.
\newblock K-shell photodetachment of {L}i$^-$: experiment and theory.
\newblock {\em Phys. Rev. Lett.} {\bf 2001}, {\em 87},~253002.
\newblock
  doi:{\changeurlcolor{black}\href{https://doi.org/10.1103/PhysRevLett.87.253002}{\detokenize{10.1103/PhysRevLett.87.253002}}}.

\bibitem[Scully \em{et~al.}(2006)Scully, {\' A}lvarez, Cisneros, Emmons,
  Gharaibeh, Leitner, Lubell, M\"{u}ller, Phaneuf, P\"{u}ttner, Schlachter,
  Schippers, and McLaughlin]{Scully2006a}
Scully, S.W.J.; {\' A}lvarez, I.; Cisneros, C.; Emmons, E.D.; Gharaibeh, M.F.;
  Leitner, D.; Lubell, M.S.; M\"{u}ller, A.; Phaneuf, R.A.; P\"{u}ttner, R.;
  Schlachter, A.S.; Schippers, S.; McLaughlin, B.M.
\newblock Doubly excited resonances in the photoionization spectrum of
  {Li$^+$}: experiment and theory.
\newblock {\em J. Phys. B} {\bf 2006}, {\em 39},~3957--3968.
\newblock
  doi:{\changeurlcolor{black}\href{https://doi.org/10.1088/0953-4075/39/18/024}{\detokenize{10.1088/0953-4075/39/18/024}}}.

\bibitem[Berrah \em{et~al.}(2007)Berrah, Bilodeau, Dumitriu, Bozek, Gibson,
  Walter, Ackerman, Zatsarinny, and Gorczyca]{Berrah2007a}
Berrah, N.; Bilodeau, R.C.; Dumitriu, I.; Bozek, J.D.; Gibson, N.D.; Walter,
  C.W.; Ackerman, G.D.; Zatsarinny, O.; Gorczyca, T.W.
\newblock Shape resonances in the absolute {K}-shell photodetachment of
  {B$^-$}.
\newblock {\em Phys. Rev. A} {\bf 2007}, {\em 76},~032713.
\newblock
  doi:{\changeurlcolor{black}\href{https://doi.org/10.1103/PhysRevA.76.032713}{\detokenize{10.1103/PhysRevA.76.032713}}}.

\bibitem[M\"{u}ller \em{et~al.}(2014)M\"{u}ller, Schippers, Phaneuf, Scully,
  Aguilar, Cisneros, Gharaibeh, Schlachter, and McLaughlin]{Mueller2014a}
M\"{u}ller, A.; Schippers, S.; Phaneuf, R.A.; Scully, S.W.J.; Aguilar, A.;
  Cisneros, C.; Gharaibeh, M.F.; Schlachter, A.S.; McLaughlin, B.M.
\newblock K -shell photoionization of {B}e-like boron ({B$^+$}) ions:
  experiment and theory.
\newblock {\em J. Phys. B} {\bf 2014}, {\em 47},~135201.
\newblock
  doi:{\changeurlcolor{black}\href{https://doi.org/10.1088/0953-4075/47/13/135201}{\detokenize{10.1088/0953-4075/47/13/135201}}}.

\bibitem[M\"{u}ller \em{et~al.}(2010)M\"{u}ller, Schippers, Phaneuf, Scully,
  Aguilar, Cisneros, Gharaibeh, Schlachter, and McLaughlin]{Mueller2010}
M\"{u}ller, A.; Schippers, S.; Phaneuf, R.A.; Scully, S.W.J.; Aguilar, A.;
  Cisneros, C.; Gharaibeh, M.F.; Schlachter, A.S.; McLaughlin, B.M.
\newblock K-shell photoionization of ground-state {L}i-like boron ions
  {[B$^{2+}$]}: experiment and theory.
\newblock {\em J. Phys. B} {\bf 2010}, {\em 43},~135602.

\bibitem[Gibson \em{et~al.}(2003)Gibson, Walter, Zatsarinny, Gorczyca,
  Ackerman, Bozek, Martins, McLaughlin, and Berrah]{Gibson2003a}
Gibson, N.D.; Walter, C.W.; Zatsarinny, O.; Gorczyca, T.W.; Ackerman, G.D.;
  Bozek, J.D.; Martins, M.; McLaughlin, B.M.; Berrah, N.
\newblock K-shell photodetachment from {C$^-$}: experiment and theory.
\newblock {\em Phys. Rev. A} {\bf 2003}, {\em 67},~030703(R).
\newblock
  doi:{\changeurlcolor{black}\href{https://doi.org/10.1103/PhysRevA.67.030703}{\detokenize{10.1103/PhysRevA.67.030703}}}.

\bibitem[Walter \em{et~al.}(2006)Walter, Gibson, Bilodeau, Berrah, Bozek,
  Ackerman, and Aguilar]{Walter2006a}
Walter, C.W.; Gibson, N.D.; Bilodeau, R.C.; Berrah, N.; Bozek, J.D.; Ackerman,
  G.D.; Aguilar, A.
\newblock Shape resonance in {K}-shell photodetachment from {C$^-$}.
\newblock {\em Phys. Rev. A} {\bf 2006}, {\em 73},~062702.
\newblock
  doi:{\changeurlcolor{black}\href{https://doi.org/10.1103/PhysRevA.73.062702}{\detokenize{10.1103/PhysRevA.73.062702}}}.

\bibitem[Perry-Sassmannshausen \em{et~al.}(2020)Perry-Sassmannshausen, Buhr,
  {Borovik Jr.}, Martins, Reinwardt, Ricz, Stock, Trinter, M\"uller, Fritzsche,
  and Schippers]{Perry-Sassmannshausen2020}
Perry-Sassmannshausen, A.; Buhr, T.; {Borovik Jr.}, A.; Martins, M.; Reinwardt,
  S.; Ricz, S.; Stock, S.O.; Trinter, F.; M\"uller, A.; Fritzsche, S.;
  Schippers, S.
\newblock Multiple photodetachment of carbon anions via single and double
  core-hole creation.
\newblock {\em Phys. Rev. Lett.} {\bf 2020}, {\em 124},~083203.
\newblock
  doi:{\changeurlcolor{black}\href{https://doi.org/10.1103/PhysRevLett.124.083203}{\detokenize{10.1103/PhysRevLett.124.083203}}}.

\bibitem[Schlachter \em{et~al.}(2004)Schlachter, Sant{'}Anna, Covington,
  Aguilar, Gharaibeh, Emmons, Scully, Phaneuf, Hinojosa, \'{A}lvarez, Cisneros,
  M\"{u}ller, and McLaughlin]{Schlachter2004a}
Schlachter, A.S.; Sant{'}Anna, M.M.; Covington, A.M.; Aguilar, A.; Gharaibeh,
  M.F.; Emmons, E.D.; Scully, S.W.J.; Phaneuf, R.A.; Hinojosa, G.; \'{A}lvarez,
  I.; Cisneros, C.; M\"{u}ller, A.; McLaughlin, B.M.
\newblock Lifetime of a {K}-shell vacancy in atomic carbon created by {$1s \to
  2p$} photoexcitation of {C$^{+}$}.
\newblock {\em J. Phys. B} {\bf 2004}, {\em 37},~L103--L109.
\newblock
  doi:{\changeurlcolor{black}\href{https://doi.org/10.1088/0953-4075/37/5/L03}{\detokenize{10.1088/0953-4075/37/5/L03}}}.

\bibitem[M\"{u}ller \em{et~al.}(2015)M\"{u}ller, Borovik, Buhr, Hellhund,
  Holste, Kilcoyne, Klumpp, Martins, Ricz, Viefhaus, and
  Schippers]{Mueller2015a}
M\"{u}ller, A.; Borovik, Jr., A.; Buhr, T.; Hellhund, J.; Holste, K.; Kilcoyne,
  A.L.D.; Klumpp, S.; Martins, M.; Ricz, S.; Viefhaus, J.; Schippers, S.
\newblock Observation of a four-electron {A}uger process in near-{K}-edge
  photoionization of singly charged carbon ions.
\newblock {\em Phys. Rev. Lett.} {\bf 2015}, {\em 114},~013002.
\newblock
  doi:{\changeurlcolor{black}\href{https://doi.org/10.1103/PhysRevLett.114.013002}{\detokenize{10.1103/PhysRevLett.114.013002}}}.

\bibitem[M\"{u}ller \em{et~al.}(2018)M\"{u}ller, Borovik, Buhr, Hellhund,
  Holste, Kilcoyne, Klumpp, Martins, Ricz, Viefhaus, and
  Schippers]{Mueller2018}
M\"{u}ller, A.; Borovik, A.; Buhr, T.; Hellhund, J.; Holste, K.; Kilcoyne,
  A.L.D.; Klumpp, S.; Martins, M.; Ricz, S.; Viefhaus, J.; Schippers, S.
\newblock Near-{K}-edge single, double, and triple photoionization of {C$^{+}$}
  ions.
\newblock {\em Phys. Rev. A} {\bf 2018}, {\em 97},~013409.
\newblock
  doi:{\changeurlcolor{black}\href{https://doi.org/10.1103/PhysRevA.97.013409}{\detokenize{10.1103/PhysRevA.97.013409}}}.

\bibitem[Scully \em{et~al.}(2005)Scully, Aguilar, Emmons, Phaneuf, Halka,
  Leitner, Levin, Lubell, P\"{u}ttner, Schlachter, Covington, Schippers,
  M\"{u}ller, and McLaughlin]{Scully2005b}
Scully, S.W.J.; Aguilar, A.; Emmons, E.D.; Phaneuf, R.A.; Halka, M.; Leitner,
  D.; Levin, J.C.; Lubell, M.S.; P\"{u}ttner, R.; Schlachter, A.S.; Covington,
  A.M.; Schippers, S.; M\"{u}ller, A.; McLaughlin, B.M.
\newblock K-shell photoionization of {B}e-like carbon ions: experiment and
  theory for {C$^{2+}$}.
\newblock {\em J. Phys. B} {\bf 2005}, {\em 38},~1967--1975.
\newblock
  doi:{\changeurlcolor{black}\href{https://doi.org/10.1088/0953-4075/38/12/011}{\detokenize{10.1088/0953-4075/38/12/011}}}.

\bibitem[M\"{u}ller \em{et~al.}(2009)M\"{u}ller, Schippers, Phaneuf, Scully,
  Aguilar, Covington, \'{A}lvarez, Cisneros, Emmons, Gharaibeh, Hinojosa,
  Schlachter, and McLaughlin]{Mueller2009a}
M\"{u}ller, A.; Schippers, S.; Phaneuf, R.A.; Scully, S.W.J.; Aguilar, A.;
  Covington, A.M.; \'{A}lvarez, I.; Cisneros, C.; Emmons, E.D.; Gharaibeh,
  M.F.; Hinojosa, G.; Schlachter, A.S.; McLaughlin, B.M.
\newblock K-shell photoionization of li-like carbon ions {[C$^{3+}$]}:
  experiment, theory and comparison with time-reversed photorecombination.
\newblock {\em J. Phys. B} {\bf 2009}, {\em 42},~235602.
\newblock
  doi:{\changeurlcolor{black}\href{https://doi.org/10.1088/0953-4075/42/23/235602}{\detokenize{10.1088/0953-4075/42/23/235602}}}.

\bibitem[Gharaibeh \em{et~al.}(2011)Gharaibeh, Bizau, Cubaynes, Guilbaud,
  El~Hassan, Al~Shorman, Miron, Nicolas, Robert, Blancard, and
  McLaughlin]{Gharaibeh2011a}
Gharaibeh, M.F.; Bizau, J.M.; Cubaynes, D.; Guilbaud, S.; El~Hassan, N.;
  Al~Shorman, M.M.; Miron, C.; Nicolas, C.; Robert, E.; Blancard, C.;
  McLaughlin, B.M.
\newblock {K}-shell photoionization of singly ionized atomic nitrogen:
  experiment and theory.
\newblock {\em J. Phys. B} {\bf 2011}, {\em 44},~175208.
\newblock
  doi:{\changeurlcolor{black}\href{https://doi.org/10.1088/0953-4075/44/17/175208}{\detokenize{10.1088/0953-4075/44/17/175208}}}.

\bibitem[Bari \em{et~al.}(2019)Bari, Inhester, Schubert, Mertens, Schunck,
  D{\"{o}}rner, Deinert, Schwob, Schippers, M{\"{u}}ller, Klumpp, and
  Martins]{Bari2019}
Bari, S.; Inhester, L.; Schubert, K.; Mertens, K.; Schunck, J.O.; D{\"{o}}rner,
  S.; Deinert, S.; Schwob, L.; Schippers, S.; M{\"{u}}ller, A.; Klumpp, S.;
  Martins, M.
\newblock Inner-shell X-ray absorption spectra of the cationic series
  {NH$_y^+$} ({$y$} = 0--3).
\newblock {\em Phys. Chem. Chem. Phys.} {\bf 2019}, {\em 21},~16505--16514.
\newblock
  doi:{\changeurlcolor{black}\href{https://doi.org/10.1039/C9CP02864A}{\detokenize{10.1039/C9CP02864A}}}.

\bibitem[McLaughlin \em{et~al.}(2020)McLaughlin, Mosnier, Kennedy, Sokell,
  Bizau, Cubaynes, Guilbaud, and Carniato]{McLaughlin2020}
McLaughlin, B.M.; Mosnier, J.P.; Kennedy, E.T.; Sokell, E.; Bizau, J.M.;
  Cubaynes, D.; Guilbaud, S.; Carniato, S.
\newblock K-shell Photoionization of the {N$^+$}, {NH$^+$} and {NH$_2^+$} ions.
\newblock {\em J. Phys.: Conf. Ser.} {\bf 2020}, {\em 1412},~142007.
\newblock
  doi:{\changeurlcolor{black}\href{https://doi.org/10.1088/1742-6596/1412/14/142007}{\detokenize{10.1088/1742-6596/1412/14/142007}}}.

\bibitem[Gharaibeh \em{et~al.}(2014)Gharaibeh, El~Hassan, Shorman, Bizau,
  Cubaynes, Guilbaud, Sakho, Blancard, and McLaughlin]{Gharaibeh2014}
Gharaibeh, M.; El~Hassan, N.; Shorman, M.A.; Bizau, J.; Cubaynes, D.; Guilbaud,
  S.; Sakho, I.; Blancard, C.; McLaughlin, B.
\newblock K-shell photoionization of {B}-like atomic nitrogen ions: experiment
  and theory.
\newblock {\em J. Phys. B} {\bf 2014}, {\em 47},~065201.
\newblock
  doi:{\changeurlcolor{black}\href{https://doi.org/10.1088/0953-4075/47/6/065201}{\detokenize{10.1088/0953-4075/47/6/065201}}}.

\bibitem[Al~Shorman \em{et~al.}(2013)Al~Shorman, Gharaibeh, Bizau, Cubaynes,
  Guilbaud, El~Hassan, Miron, Nicolas, Robert, Sakho, Blancard, and
  McLaughlin]{AlShorman2013}
Al~Shorman, M.M.; Gharaibeh, M.F.; Bizau, J.M.; Cubaynes, D.; Guilbaud, S.;
  El~Hassan, N.; Miron, C.; Nicolas, C.; Robert, E.; Sakho, I.; Blancard, C.;
  McLaughlin, B.M.
\newblock K-shell photoionization of {B}e-like and {L}i-like ions of atomic
  nitrogen: experiment and theory.
\newblock {\em J. Phys. B} {\bf 2013}, {\em 46},~195701.
\newblock
  doi:{\changeurlcolor{black}\href{https://doi.org/10.1088/0953-4075/46/19/195701}{\detokenize{10.1088/0953-4075/46/19/195701}}}.

\bibitem[Gibson \em{et~al.}(2012)Gibson, Bilodeau, Walter, Hanstorp, Aguilar,
  Berrah, Matyas, Li, Alton, and Lou]{Gibson2012}
Gibson, N.D.; Bilodeau, R.C.; Walter, C.W.; Hanstorp, D.; Aguilar, A.; Berrah,
  N.; Matyas, D.J.; Li, Y.G.; Alton, R.M.; Lou, S.E.
\newblock K-shell photodetachment from {O$^-$}.
\newblock {\em J. Phys.: Conf. Ser.} {\bf 2012}, {\em 388},~022102.
\newblock
  doi:{\changeurlcolor{black}\href{https://doi.org/10.1088/1742-6596/388/2/022102}{\detokenize{10.1088/1742-6596/388/2/022102}}}.

\bibitem[Schippers \em{et~al.}(2016)Schippers, Beerwerth, Abrok, Bari, Buhr,
  Martins, Ricz, Viefhaus, Fritzsche, and M\"{u}ller]{Schippers2016a}
Schippers, S.; Beerwerth, R.; Abrok, L.; Bari, S.; Buhr, T.; Martins, M.; Ricz,
  S.; Viefhaus, J.; Fritzsche, S.; M\"{u}ller, A.
\newblock Prominent role of multielectron processes in {K}-shell double and
  triple photodetachment of oxygen anions.
\newblock {\em Phys. Rev. A} {\bf 2016}, {\em 94},~041401.
\newblock
  doi:{\changeurlcolor{black}\href{https://doi.org/10.1103/PhysRevA.94.041401}{\detokenize{10.1103/PhysRevA.94.041401}}}.

\bibitem[Kawatsura \em{et~al.}(2002)Kawatsura, Yamaoka, Oura, Hayaishi,
  Sekioka, Agui, Yoshigoe, and Koike]{Kawatsura2002a}
Kawatsura, K.; Yamaoka, H.; Oura, M.; Hayaishi, T.; Sekioka, T.; Agui, A.;
  Yoshigoe, A.; Koike, F.
\newblock The {$1s-2p$} resonance photoionization measurement of {O$^+$} ions
  in comparison with an isoelectronic species {Ne$^{3+}$}.
\newblock {\em J. Phys. B} {\bf 2002}, {\em 35},~4147--4153.

\bibitem[Bizau \em{et~al.}(2015)Bizau, Cubaynes, Guilbaud, Al~Shorman,
  Gharaibeh, Ababneh, Blancard, and McLaughlin]{Bizau2015}
Bizau, J.M.; Cubaynes, D.; Guilbaud, S.; Al~Shorman, M.M.; Gharaibeh, M.F.;
  Ababneh, I.Q.; Blancard, C.; McLaughlin, B.M.
\newblock {K}-shell photoionization of {O$^{+}$} and {O$^{2+}$} ions:
  experiment and theory.
\newblock {\em Phys. Rev. A} {\bf 2015}, {\em 92},~023401.
\newblock
  doi:{\changeurlcolor{black}\href{https://doi.org/10.1103/PhysRevA.92.023401}{\detokenize{10.1103/PhysRevA.92.023401}}}.

\bibitem[McLaughlin \em{et~al.}(2014)McLaughlin, Bizau, Cubaynes, Shorman,
  Guilbaud, Sakho, Blancard, and Gharaibeh]{McLaughlin2014}
McLaughlin, B.; Bizau, J.; Cubaynes, D.; Shorman, M.A.; Guilbaud, S.; Sakho,
  I.; Blancard, C.; Gharaibeh, M.
\newblock K-shell photoionization of {B}-like oxygen {(O$^{3+}$)} ions:
  experiment and theory.
\newblock {\em J. Phys. B} {\bf 2014}, {\em 47},~115201.
\newblock
  doi:{\changeurlcolor{black}\href{https://doi.org/10.1088/0953-4075/47/11/115201}{\detokenize{10.1088/0953-4075/47/11/115201}}}.

\bibitem[McLaughlin \em{et~al.}(2017)McLaughlin, Bizau, Cubaynes, Guilbaud,
  Douix, Shorman, Ghazaly, Sakho, and Gharaibeh]{McLaughlin2017}
McLaughlin, B.M.; Bizau, J.M.; Cubaynes, D.; Guilbaud, S.; Douix, S.; Shorman,
  M.M.A.; Ghazaly, M.O.A.E.; Sakho, I.; Gharaibeh, M.F.
\newblock {K-shell photoionization of O$^{4 +}$ and O$^{5 +}$ ions: experiment
  and theory}.
\newblock {\em Mon. Not. R. Astron. Soc.} {\bf 2017}, {\em 465},~4690--4702.
\newblock
  doi:{\changeurlcolor{black}\href{https://doi.org/10.1093/mnras/stw2998}{\detokenize{10.1093/mnras/stw2998}}}.

\bibitem[M\"{u}ller \em{et~al.}(2018)M\"{u}ller, {Borovik Jr.}, Bari, Buhr,
  Holste, Martins, Perry-Sassmannshausen, Phaneuf, Reinwardt, Ricz, Schubert,
  and Schippers]{Mueller2018b}
M\"{u}ller, A.; {Borovik Jr.}, A.; Bari, S.; Buhr, T.; Holste, K.; Martins, M.;
  Perry-Sassmannshausen, A.; Phaneuf, R.A.; Reinwardt, S.; Ricz, S.; Schubert,
  K.; Schippers, S.
\newblock Near-{K}-edge double and triple detachment of the {F$^-$} negative
  ion: observation of direct two-electron ejection by a single photon.
\newblock {\em Phys. Rev. Lett.} {\bf 2018}, {\em 120},~133202.
\newblock
  doi:{\changeurlcolor{black}\href{https://doi.org/10.1103/PhysRevLett.120.133202}{\detokenize{10.1103/PhysRevLett.120.133202}}}.

\bibitem[Oura \em{et~al.}(2001)Oura, Yamaoka, Kawatsura, Kimata, Hayaishi,
  Takahashi, Koizumi, Sekioka, Terasawa, Itoh, Awaya, Yokoya, Agui, Yoshigoe,
  and Saitoh]{Oura2001}
Oura, M.; Yamaoka, H.; Kawatsura, K.; Kimata, J.; Hayaishi, T.; Takahashi, T.;
  Koizumi, T.; Sekioka, T.; Terasawa, M.; Itoh, Y.; Awaya, Y.; Yokoya, A.;
  Agui, A.; Yoshigoe, A.; Saitoh, Y.
\newblock Photoionization of {Ne$^{3+}$} ions in the region of the {$1s2p$}
  autoionizing resonance.
\newblock {\em Phys. Rev. A} {\bf 2001}, {\em 63},~014704.

\bibitem[Buhr \em{et~al.}(2020)Buhr, Stock, Perry-Sassmannshausen, Reinwardt,
  Martins, Ricz, M{\"{u}}ller, Fritzsche, and Schippers]{Buhr2020}
Buhr, T.; Stock, S.O.; Perry-Sassmannshausen, A.; Reinwardt, S.; Martins, M.;
  Ricz, S.; M{\"{u}}ller, A.; Fritzsche, S.; Schippers, S.
\newblock Photoionization of low-charged silicon ions.
\newblock {\em J. Phys.: Conf. Ser.} {\bf 2020}, {\em 1412},~152024.
\newblock
  doi:{\changeurlcolor{black}\href{https://doi.org/10.1088/1742-6596/1412/15/152024}{\detokenize{10.1088/1742-6596/1412/15/152024}}}.

\bibitem[M\"{u}ller(2015)]{Mueller2015}
M\"{u}ller, A.
\newblock Precision studies of deep-inner-shell photoabsorption by atomic ions.
\newblock {\em Phys. Scr.} {\bf 2015}, {\em 90},~054004.
\newblock
  doi:{\changeurlcolor{black}\href{https://doi.org/10.1088/0031-8949/90/5/054004}{\detokenize{10.1088/0031-8949/90/5/054004}}}.

\bibitem[Gorczyca(2000)]{Gorczyca2000a}
Gorczyca, T.W.
\newblock Auger decay of the photoexcited {$1s^{-1}np$} {R}ydberg series in
  neon.
\newblock {\em Phys. Rev. A} {\bf 2000}, {\em 61},~024702.
\newblock
  doi:{\changeurlcolor{black}\href{https://doi.org/10.1103/PhysRevA.61.024702}{\detokenize{10.1103/PhysRevA.61.024702}}}.

\bibitem[Gatuzz \em{et~al.}(2015)Gatuzz, Garcia, Kallman, Mendoza, and
  Gorczyca]{Gatuzz2015}
Gatuzz, E.; Garcia, J.; Kallman, T.R.; Mendoza, C.; Gorczyca, T.W.
\newblock {ISM}abs: a comprehensive x-ray absorption model for the interstellar
  medium.
\newblock {\em Astrophys. J.} {\bf 2015}, {\em 800},~29.
\newblock
  doi:{\changeurlcolor{black}\href{https://doi.org/10.1088/0004-637X/800/1/29}{\detokenize{10.1088/0004-637X/800/1/29}}}.

\bibitem[Witthoeft \em{et~al.}(2009)Witthoeft, Bautista, Mendoza, Kallman,
  Palmeri, and Quinet]{Witthoeft2009}
Witthoeft, M.C.; Bautista, M.A.; Mendoza, C.; Kallman, T.R.; Palmeri, P.;
  Quinet, P.
\newblock K-shell photoionization and photoabsorption of {N}e, {M}g, {S}i, {S},
  {A}r, and {C}a.
\newblock {\em Astrophys. J. Suppl. Ser.} {\bf 2009}, {\em 182},~127--130.
\newblock
  doi:{\changeurlcolor{black}\href{https://doi.org/10.1088/0067-0049/182/1/127}{\detokenize{10.1088/0067-0049/182/1/127}}}.

\bibitem[Liao \em{et~al.}(2013)Liao, Zhang, and Yao]{Liao2013}
Liao, J.Y.; Zhang, S.N.; Yao, Y.
\newblock Wavelength measurements of {K} transitions of oxygen, neon, and
  magnesium with x-ray absorption lines.
\newblock {\em Astrophys. J.} {\bf 2013}, {\em 774},~116.
\newblock
  doi:{\changeurlcolor{black}\href{https://doi.org/10.1088/0004-637x/774/2/116}{\detokenize{10.1088/0004-637x/774/2/116}}}.

\bibitem[Landi and Lepri(2015)]{Landi2015}
Landi, E.; Lepri, S.T.
\newblock Photoionization in the {S}olar wind.
\newblock {\em Astrophys. J.} {\bf 2015}, {\em 812},~L28.
\newblock
  doi:{\changeurlcolor{black}\href{https://doi.org/10.1088/2041-8205/812/2/l28}{\detokenize{10.1088/2041-8205/812/2/l28}}}.

\bibitem[Larsson \em{et~al.}(2012)Larsson, Geppert, and Nyman]{Larsson2012}
Larsson, M.; Geppert, W.D.; Nyman, G.
\newblock Ion chemistry in space.
\newblock {\em Rep. Prog. Phys.} {\bf 2012}, {\em 75},~066901.
\newblock
  doi:{\changeurlcolor{black}\href{https://doi.org/10.1088/0034-4885/75/6/066901}{\detokenize{10.1088/0034-4885/75/6/066901}}}.

\bibitem[Tielens(2013)]{Tielens2013}
Tielens, A.G.G.M.
\newblock The molecular universe.
\newblock {\em Rev. Mod. Phys.} {\bf 2013}, {\em 85},~1021--1081.
\newblock
  doi:{\changeurlcolor{black}\href{https://doi.org/10.1103/RevModPhys.85.1021}{\detokenize{10.1103/RevModPhys.85.1021}}}.

\bibitem[Millar \em{et~al.}(2017)Millar, Walsh, and Field]{Millar2017}
Millar, T.J.; Walsh, C.; Field, T.A.
\newblock Negative ions in space.
\newblock {\em Chem. Rev.} {\bf 2017}, {\em 117},~1765--1795.
\newblock
  doi:{\changeurlcolor{black}\href{https://doi.org/10.1021/acs.chemrev.6b00480}{\detokenize{10.1021/acs.chemrev.6b00480}}}.

\bibitem[Mosnier \em{et~al.}(2016)Mosnier, Kennedy, van Kampen, Cubaynes,
  Guilbaud, Sisourat, Puglisi, Carniato, and Bizau]{Mosnier2016}
Mosnier, J.P.; Kennedy, E.T.; van Kampen, P.; Cubaynes, D.; Guilbaud, S.;
  Sisourat, N.; Puglisi, A.; Carniato, S.; Bizau, J.M.
\newblock Inner-shell photoexcitations as probes of the molecular ions
  {CH$^{+}$}, {OH$^{+}$}, and {SiH$^{+}$}: measurements and theory.
\newblock {\em Phys. Rev. A} {\bf 2016}, {\em 93},~061401.
\newblock
  doi:{\changeurlcolor{black}\href{https://doi.org/10.1103/PhysRevA.93.061401}{\detokenize{10.1103/PhysRevA.93.061401}}}.

\bibitem[Hellhund \em{et~al.}(2015)Hellhund, Borovik~Jr., Holste, Klumpp,
  Martins, Ricz, Schippers, and M\"{u}ller]{Hellhund2015}
Hellhund, J.; Borovik~Jr., A.; Holste, K.; Klumpp, S.; Martins, M.; Ricz, S.;
  Schippers, S.; M\"{u}ller, A.
\newblock Photoionization and photofragmentation of multiply charged
  {L}u{$_3$}N@{C}{$_{80}$} ions.
\newblock {\em Phys. Rev. A} {\bf 2015}, {\em 92},~013413.
\newblock
  doi:{\changeurlcolor{black}\href{https://doi.org/10.1103/PhysRevA.92.013413}{\detokenize{10.1103/PhysRevA.92.013413}}}.

\bibitem[M\"uller \em{et~al.}(2019)M\"uller, Martins, Kilcoyne, Phaneuf,
  Hellhund, Borovik, Holste, Bari, Buhr, Klumpp, Perry-Sassmannshausen,
  Reinwardt, Ricz, Schubert, and Schippers]{Mueller2019}
M\"uller, A.; Martins, M.; Kilcoyne, A.L.D.; Phaneuf, R.A.; Hellhund, J.;
  Borovik, A.; Holste, K.; Bari, S.; Buhr, T.; Klumpp, S.;
  Perry-Sassmannshausen, A.; Reinwardt, S.; Ricz, S.; Schubert, K.; Schippers,
  S.
\newblock Photoionization and photofragmentation of singly charged positive and
  negative {Sc$_{3}$N@C$_{80}$} endohedral fullerene ions.
\newblock {\em Phys. Rev. A} {\bf 2019}, {\em 99},~063401.
\newblock
  doi:{\changeurlcolor{black}\href{https://doi.org/10.1103/PhysRevA.99.063401}{\detokenize{10.1103/PhysRevA.99.063401}}}.

\bibitem[Voulot \em{et~al.}(2000)Voulot, Gillen, Thompson, Gilbody, McCullough,
  Errea, Macias, Mendez, and Riera]{Voulot2000}
Voulot, D.; Gillen, D.R.; Thompson, W.R.; Gilbody, H.B.; McCullough, R.W.;
  Errea, L.; Macias, A.; Mendez, L.; Riera, A.
\newblock First studies of state-selective electron capture in collisions of
  state-prepared ions with atomic hydrogen; the case of {C}$^{2+}$--{H}(1s).
\newblock {\em J. Phys. B} {\bf 2000}, {\em 33},~L187--L192.
\newblock
  doi:{\changeurlcolor{black}\href{https://doi.org/10.1088/0953-4075/33/5/106}{\detokenize{10.1088/0953-4075/33/5/106}}}.

\bibitem[Covington \em{et~al.}(2001)Covington, Aguilar, Covington, Gharaibeh,
  Shirley, Phaneuf, \'{A}lvarez, Cisneros, Hinojosa, Bozek, Dominguez,
  Sant{'}Anna, Schlachter, Berrah, Nahar, and McLaughlin]{Covington2001a}
Covington, A.M.; Aguilar, A.; Covington, I.R.; Gharaibeh, M.; Shirley, C.A.;
  Phaneuf, R.A.; \'{A}lvarez, I.; Cisneros, C.; Hinojosa, G.; Bozek, J.D.;
  Dominguez, I.; Sant{'}Anna, M.M.; Schlachter, A.S.; Berrah, N.; Nahar, S.N.;
  McLaughlin, B.M.
\newblock Photoionization of metastable {O$^+$} ions: experiment and theory.
\newblock {\em Phys. Rev. Lett.} {\bf 2001}, {\em 87},~243002.
\newblock
  doi:{\changeurlcolor{black}\href{https://doi.org/10.1103/PhysRevLett.87.243002}{\detokenize{10.1103/PhysRevLett.87.243002}}}.

\bibitem[Benis \em{et~al.}(2018)Benis, Madesis, Laoutaris, Nanos, and
  Zouros]{Benis2018}
Benis, E.P.; Madesis, I.; Laoutaris, A.; Nanos, S.; Zouros, T.J.M.
\newblock Mixed-state ionic beams: an effective tool for collision dynamics
  investigations.
\newblock {\em Atoms} {\bf 2018}, {\em 6}.
\newblock
  doi:{\changeurlcolor{black}\href{https://doi.org/10.3390/atoms6040066}{\detokenize{10.3390/atoms6040066}}}.

\bibitem[Schippers \em{et~al.}(2002)Schippers, M\"{u}ller, Ricz, Bannister,
  Dunn, Bozek, Schlachter, Hinojosa, Cisneros, Aguilar, Covington, Gharaibeh,
  and Phaneuf]{Schippers2002b}
Schippers, S.; M\"{u}ller, A.; Ricz, S.; Bannister, M.E.; Dunn, G.H.; Bozek,
  J.D.; Schlachter, A.S.; Hinojosa, G.; Cisneros, C.; Aguilar, A.; Covington,
  A.M.; Gharaibeh, M.F.; Phaneuf, R.A.
\newblock Experimental link of photoionization of {Sc$^{2+}$} to
  photorecombination of {Sc$^{3+}$}: an application of detailed balance in a
  unique atomic system.
\newblock {\em Phys. Rev. Lett.} {\bf 2002}, {\em 89},~193002.
\newblock
  doi:{\changeurlcolor{black}\href{https://doi.org/10.1103/PhysRevLett.89.193002}{\detokenize{10.1103/PhysRevLett.89.193002}}}.

\bibitem[Wolf \em{et~al.}(1991)Wolf, Berger, Bock, Habs, Hochadel, Kilgus,
  Neureither, Schramm, Schwalm, Szmola, M\"{u}ller, Wagner, and
  Schuch]{Wolf1991}
Wolf, A.; Berger, J.; Bock, M.; Habs, D.; Hochadel, B.; Kilgus, G.; Neureither,
  G.; Schramm, U.; Schwalm, D.; Szmola, E.; M\"{u}ller, A.; Wagner, M.; Schuch,
  R.
\newblock Experiments with highly-charged ions in the storage ring {TSR}.
\newblock {\em Z. Phys. D} {\bf 1991}, {\em 21},~S69--S75.
\newblock
  doi:{\changeurlcolor{black}\href{https://doi.org/10.1007/BF01426253}{\detokenize{10.1007/BF01426253}}}.

\bibitem[Andersen(2004)]{Andersen2004b}
Andersen, T.
\newblock Atomic negative ions: structure, dynamics and collisions.
\newblock {\em Phys. Reports} {\bf 2004}, {\em 394},~157--313.
\newblock
  doi:{\changeurlcolor{black}\href{https://doi.org/10.1016/j.physrep.2004.01.001}{\detokenize{10.1016/j.physrep.2004.01.001}}}.

\bibitem[Schippers(2018)]{Schippers2018}
Schippers, S.
\newblock Analytical expression for the convolution of a {F}ano line profile
  with a gaussian.
\newblock {\em J. Quant. Spectrosc. Radiat. Transfer} {\bf 2018}, {\em
  219},~33--36.
\newblock
  doi:{\changeurlcolor{black}\href{https://doi.org/10.1016/j.jqsrt.2018.08.003}{\detokenize{10.1016/j.jqsrt.2018.08.003}}}.

\bibitem[Lindroth and Argenti(2012)]{Lindroth2012}
Lindroth, E.; Argenti, L.
\newblock Atomic resonance states and their role in charge-changing processes.
\newblock {\em Adv. Quantum Chem.} {\bf 2012}, {\em 63},~247.
\newblock
  doi:{\changeurlcolor{black}\href{https://doi.org/10.1016/B978-0-12-397009-1.00005-9}{\detokenize{10.1016/B978-0-12-397009-1.00005-9}}}.

\bibitem[Indelicato \em{et~al.}(1987)Indelicato, Gorveix, and
  Desclaux]{Indelicato1987}
Indelicato, P.; Gorveix, O.; Desclaux, J.P.
\newblock Multiconfigurational {D}irac-{F}ock studies of two-electron ions.
  {II}. Radiative corrections and comparison with experiment.
\newblock {\em J. Phys. B} {\bf 1987}, {\em 20},~651.
\newblock
  doi:{\changeurlcolor{black}\href{https://doi.org/10.1088/0022-3700/20/4/007}{\detokenize{10.1088/0022-3700/20/4/007}}}.

\bibitem[Zeegers \em{et~al.}(2019)Zeegers, Costantini, Rogantini, de~Vries,
  Mutschke, Mohr, de~Groot, and Tielens]{Zeegers2019}
Zeegers, S.T.; Costantini, E.; Rogantini, D.; de~Vries, C.P.; Mutschke, H.;
  Mohr, P.; de~Groot, F.; Tielens, A.G.G.M.
\newblock Dust absorption and scattering in the silicon {K}-edge.
\newblock {\em Astron. Astrophys.} {\bf 2019}, {\em 627},~A16.
\newblock
  doi:{\changeurlcolor{black}\href{https://doi.org/10.1051/0004-6361/201935050}{\detokenize{10.1051/0004-6361/201935050}}}.

\bibitem[Schulz \em{et~al.}(2016)Schulz, Corrales, and Canizares]{Schulz2016}
Schulz, N.S.; Corrales, L.; Canizares, C.R.
\newblock Si {K} edge structure and variability in {G}alactic x-ray binaries.
\newblock {\em Astrophys. J.} {\bf 2016}, {\em 827},~49.
\newblock
  doi:{\changeurlcolor{black}\href{https://doi.org/10.3847/0004-637x/827/1/49}{\detokenize{10.3847/0004-637x/827/1/49}}}.

\bibitem[Witthoeft \em{et~al.}(2011)Witthoeft, Garc\'{i}a, Kallman, Bautista,
  Mendoza, Palmeri, and Quinet]{Witthoeft2011}
Witthoeft, M.C.; Garc\'{i}a, J.; Kallman, T.R.; Bautista, M.A.; Mendoza, C.;
  Palmeri, P.; Quinet, P.
\newblock K-shell photoionization of {N}a-like to {C}l-like ions of {M}g, {S}i,
  {S}, {A}r, and {C}a.
\newblock {\em Astrophys. J. Suppl. Ser.} {\bf 2011}, {\em 1992},~7.
\newblock arXiv:1010.3734v1 [physics.atom-ph],
  doi:{\changeurlcolor{black}\href{https://doi.org/10.1088/0067-0049/192/1/7}{\detokenize{10.1088/0067-0049/192/1/7}}}.

\bibitem[Ku\v{c}as \em{et~al.}(2012)Ku\v{c}as, Karazija, and
  Momkauskait\.{e}]{Kucas2012}
Ku\v{c}as, S.; Karazija, R.; Momkauskait\.{e}, A.
\newblock Cascades after {K}-vacancy production in atoms and ions of light
  elements.
\newblock {\em Astrophys. J.} {\bf 2012}, {\em 750},~90.
\newblock
  doi:{\changeurlcolor{black}\href{https://doi.org/10.1088/0004-637X/750/2/90}{\detokenize{10.1088/0004-637X/750/2/90}}}.

\bibitem[Hasoglu and Gorczyca(2018)]{Hasoglu2018}
Hasoglu, M.F.; Gorczyca, T.W.
\newblock X-Ray Absorption by Interstellar Atomic Gases near the {K} Edges of
  {C}, {O}, {N}e, {M}g, and {S}i and the {L} Edge of {F}e.
\newblock {\em ASP Conf. Ser.} {\bf 2018}, {\em 515},~275.

\bibitem[Ueda \em{et~al.}(2019)Ueda, Sokell, Schippers, Aumayr, Sadeghpour,
  Burgd\"orfer, Lemell, Tong, Pfeifer, Calegari, Palacios, Martin, Corkum,
  Sansone, Gryzlova, Grum-Grzhimailo, Piancastelli, Weber, Steinle, Amini,
  Biegert, Berrah, Kukk, Santra, M\"uller, Dowek, Lucchese, McCurdy, Bolognesi,
  Avaldi, Jahnke, Sch\"offler, D\"orner, Mairesse, Nahon, Smirnova,
  Schlath\"olter, Campbell, Rost, Meyer, and Tanaka]{Ueda2019}
Ueda, K.; Sokell, E.; Schippers, S.; Aumayr, F.; Sadeghpour, H.; Burgd\"orfer,
  J.; Lemell, C.; Tong, X.M.; Pfeifer, T.; Calegari, F.; Palacios, A.; Martin,
  F.; Corkum, P.; Sansone, G.; Gryzlova, E.V.; Grum-Grzhimailo, A.N.;
  Piancastelli, M.N.; Weber, P.M.; Steinle, T.; Amini, K.; Biegert, J.; Berrah,
  N.; Kukk, E.; Santra, R.; M\"uller, A.; Dowek, D.; Lucchese, R.R.; McCurdy,
  C.W.; Bolognesi, P.; Avaldi, L.; Jahnke, T.; Sch\"offler, M.S.; D\"orner, R.;
  Mairesse, Y.; Nahon, L.; Smirnova, O.; Schlath\"olter, T.; Campbell, E.E.B.;
  Rost, J.M.; Meyer, M.; Tanaka, K.A.
\newblock Roadmap on photonic, electronic and atomic collision physics: {I.
  L}ight--matter interaction.
\newblock {\em J. Phys. B} {\bf 2019}, {\em 52},~171001.
\newblock
  doi:{\changeurlcolor{black}\href{https://doi.org/10.1088/1361-6455/ab26d7}{\detokenize{10.1088/1361-6455/ab26d7}}}.

\end{thebibliography}

\end{document}